\documentclass[10pt,conference]{IEEEtran}
\usepackage{cite}
\usepackage{amsmath,amssymb,amsfonts}
\usepackage{algorithmicx}
\usepackage{algcompatible}
\usepackage{graphicx}
\usepackage{textcomp}
\usepackage{xcolor}
\usepackage[hyphens]{url}
\usepackage{fancyhdr}
\usepackage{hyperref}
\usepackage{algorithm}
\usepackage{multicol}
\usepackage[noend]{algpseudocode}

\newcommand{\hpcayear}{2026}

\newcommand{\hpcasubmissionnumber}{1077}
\title{Cyclone: Designing Efficient and Highly Parallel
QCCD Architectural Codesigns for Fault Tolerant
Quantum Memory}
\def\hpcacameraready{} % Uncomment to build camera-ready version

\newcommand\hpcaauthors{Sahil Khan$\dagger$, Abhinav Anand$\dagger$, Kenneth R. Brown$\dagger$, and Jonathan M. Baker$\ddagger$}
\newcommand\hpcaaffiliation{Duke University$\dagger$, University of Texas at Austin$\ddagger$}
\newcommand\hpcaemail{sahil.khan@duke.edu}

\begin{document}
%%%%%%%%%%%%%%%%%%%%%%%%%%%%%%%%%%%%%
%%%%%%%%%% -- DO NOT MODIFY -- %%%%%%%%%%
%%%%%%%%%%%%%%%%%%%%%%%%%%%%%%%%%%%%%

\author{
  \ifdefined\hpcacameraready
    \IEEEauthorblockN{\hpcaauthors{}}
      \IEEEauthorblockA{
        \hpcaaffiliation{} \\
        \hpcaemail{}
      }
  \else
    \IEEEauthorblockN{\normalsize{HPCA \hpcayear{} Submission
      \textbf{\#\hpcasubmissionnumber{}}} \\
      \IEEEauthorblockA{
        Confidential Draft \\
        Do NOT Distribute!!
      }
    }
  \fi 
}

% Heading and footer for title page
\fancypagestyle{camerareadyfirstpage}{%
  \fancyhead{}
  \renewcommand{\headrulewidth}{0pt}
  \fancyhead[C]{
    \ifdefined\aeopen
    \parbox[][12mm][t]{13.5cm}{\hpcayear{} IEEE International Symposium on High-Performance Computer Architecture (HPCA)}    
    \else
      \ifdefined\aereviewed
      \parbox[][12mm][t]{13.5cm}{\hpcayear{} IEEE International Symposium on High-Performance Computer Architecture (HPCA)}
      \else
      \ifdefined\aereproduced
      \parbox[][12mm][t]{13.5cm}{\hpcayear{} IEEE International Symposium on High-Performance Computer Architecture (HPCA)}
      \else
      \parbox[][0mm][t]{13.5cm}{\hpcayear{} IEEE International Symposium on High-Performance Computer Architecture (HPCA)}
    \fi 
    \fi 
    \fi 
    \ifdefined\aeopen 
      \includegraphics[width=12mm,height=12mm]{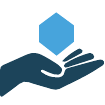}
    \fi 
    \ifdefined\aereviewed
      \includegraphics[width=12mm,height=12mm]{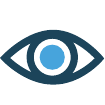}
    \fi 
    \ifdefined\aereproduced
      \includegraphics[width=12mm,height=12mm]{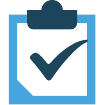}
    \fi
  }
  %\fancyfoot[L]{\hpcapubid{} \copyright \hpcayear{} IEEE}
  \fancyfoot[C]{}
}
% Heading and footer for remaining pages
\fancyhead{}
\renewcommand{\headrulewidth}{0pt}
%\fancyhead[C]{\hpcayear{} IEEE International Symposium on
% High-Performance Computer Architecture (HPCA)}

%Enables the camera ready header and footer
\ifdefined\hpcacameraready 
  \thispagestyle{camerareadyfirstpage}
  \pagestyle{empty}
\else
  \thispagestyle{plain}
  \pagestyle{plain}
\fi

\newcommand{\hpcaheight}{0mm}
\ifdefined\eaopen
\renewcommand{\hpcaheight}{12mm}
\fi

\maketitle %moved outside of hpca template file by Sahil
%%%%%%%%%%%%%%%%%%%%%%%%%%%%%%%%%%%%%%%%
%%%%%%%% -- PAPER CONTENT STARTS -- %%%%%%%%%

\begin{abstract}
Modular trapped-ion quantum computing hardware, known as Quantum Charge Coupled Devices (QCCDs) require shuttling operations in order to maintain effective all-to-all connectivity. 
Each module or trap can perform only one operation at a time, resulting in \textit{low intra-trap} parallelism, but there is no restriction on operations happening on independent traps, enabling \textit{high inter-trap} parallelism. 
Unlike their superconducting counterparts, the design space for QCCDs is relatively flexible and can be explored beyond the constraints of two-dimensional grids.
In this work, we are motivated by the opportunity to explore the QCCD design space in the context of optimizing for non-topological CSS codes.
In particular, current grid-based architectures significantly limit the performance of many promising, high-rate codes such as hypergraph product codes and bivariate bicycle codes. 
Many of these codes are highly parallelizable, meaning that with appropriate hardware layouts and matching software schedules, execution latency can be greatly reduced. 
Faster execution, in turn, reduces error accumulation from decoherence and heating, ultimately improving code performance when mapped to realistic hardware.

However, current 2D grid designs suffer from numerous trap to trap ``roadblocks", forcing serialization and destroying the inherent parallelism of these codes.
To address this, we propose \textit{Cyclone}, a circular software-hardware codesign that departs from traditional 2D grids in favor of a flexible ring topology, where ancilla qubits move in lockstep.
Cyclone eliminates roadblocks, bounds total movement, and enables high levels of parallelism, resulting in up to ~4$\times$ speedup in execution times. 
In addition to temporal efficiency, Cyclone also offers large spatial efficiency when compared to a grid codesign.
It requires fewer traps, fewer junctions, and only a constant number of Digital-to-Analog Converters (DAC), as opposed to grid architectures, where DAC count scales linearly with the number of traps.
With hypergraph product codes, Cyclone achieves up to a 2$\times$ order of magnitude improvement in logical error rate, and with bivariate bicycle codes, this improvement reaches up to a 3$\times$ in order of magnitude.
Spatially, Cyclone reduces the number of required traps and ancilla qubits by $2\times$.
The overall spacetime improvement over a standard grid is up to $\sim 20 \times$, demonstrating Cyclone as a scalable and efficient alternative to conventional 2D QCCD architectures.

\end{abstract}

\section{Introduction}
\begin{figure*}[htbp!]
    \centering
    \includegraphics[width=0.95\linewidth]{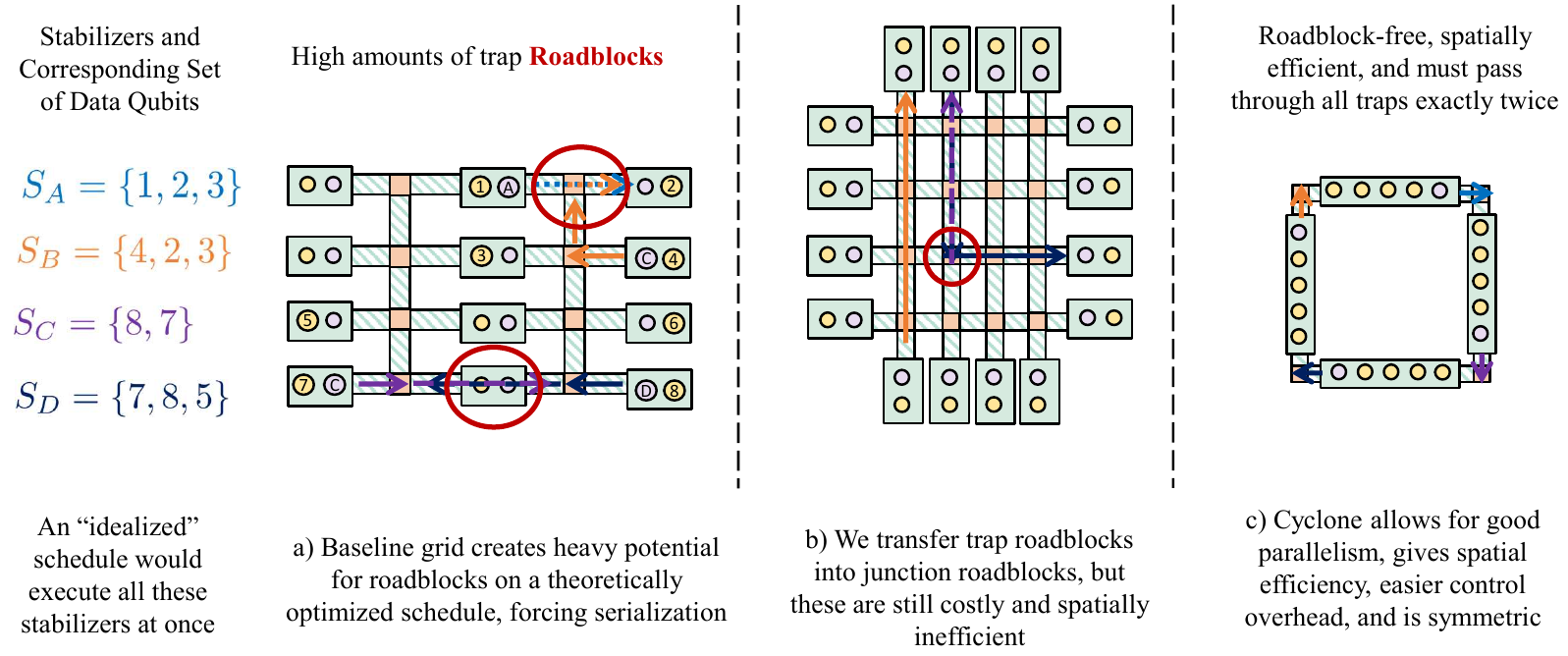}
    \caption{Left: the baseline grid architectural codesign is shown. This leads to gridlock, and large amounts of serialization on large arbitrary QEC codes with possible long-distance connections (non-topological codes). This design has both high \textit{spatial} and \textit{temporal} overhead. Middle: a mesh junction network is shown, which has high spatial requirements for junctions and bad temporal efficiency due to junction crossing overhead. Right: a symmetric roadblock-free design is shown. This design has low \textit{spatial} and \textit{temporal} overhead, with low control overhead due to its simplicity.}
    \label{fig:fig1}
\end{figure*}

Trapped-ion quantum computing systems utilizes electrodynamic traps to confine ions, with lasers used to control their quantum state and enable arbitrary quantum operations.
However, a single trap cannot scale to hundreds of ions without suffering from excessive heating rates and control overhead \cite{chrisqccdinvention, brown2016codesigningscalablequantumcomputer, murali}. 
Additionally, \textit{each trap} has \textit{low parallelism}, with most hardware only allowing one or two gates in parallel within a trap. Therefore, scaling a trapped-ion quantum computer requires a modular architecture, where traps are inter-connected via \textit{shuttling paths}, \textit{photonic links}, or a combination of both to maintain effective all-to-all connectivity across hardware \cite{musiqc, photonic_parallelism}.

In this paper, we focus on the Quantum Charge Coupled Device (QCCD) paradigm \cite{chrisqccdinvention, brown2016codesigningscalablequantumcomputer, QCCDdemonstration}, which uses a network of shuttling operations and junctions to connect multiple traps together. QCCD devices have no restrictions on parallel operations in different traps, enabling \textit{high inter-trap parallelism} \cite{ovide2025exploringoperationparallelismvs, photonic_parallelism}. Upon compilation to hardware, program qubits are mapped onto the target device, gates are scheduled, and shuttling operations are inserted, resulting in a deeper compiled circuit with a longer execution latency. Changing the hardware topology significantly influences the optimal mapping and scheduling policy for a given circuit, due to the hardware's connectivity limitations and support for parallelism. Since the shuttling network configuration, trap layout, and compilation policies (both mapping and scheduling) are tunable, there is a large opportunity to explore the design space of scalable trapped-ion architectures, as opposed to their superconducting counterparts that have more
restricted layouts, e.g. 2D grids \cite{murali, scalability_challenges}.

In this work, we explore the design space in the context of optimizing fault-tolerant quantum memory. Fault-tolerant memory is achieved by implementing a quantum error-correcting (QEC) code that uses multiple physical \textit{data qubits} to create a more error-resilient \textit{logical qubit} ~\cite{kitaev1997toriccodes, landahl2011colorcodes, shor1995qec}. The most popular family of codes are \textit{CSS stabilizer codes} \cite{gottesman1997stabilizer}, which detect errors by entangling the data qubits with ancilla qubits to measure Pauli operators, known as stabilizers. This stabilizer measurement process (commonly referred to as \textit{syndrome extraction}) is repeated multiple times to generate a matrix of classical bits, known as the full syndrome matrix. A classical decoder then analyzes the syndrome to detect the errors, and determine the appropriate correction. If the proposed correction results in the logical state being flipped, a \textit{logical error} occurs. This process is repeated indefinitely with the goal of preserving the logical state. The circuit that measures the syndrome each round is known as the \textit{syndrome extraction circuit}, which is \textbf{highly parallelizable} for many stabilizer codes. However, the syndrome extraction circuit itself can be faulty, and because decoherence impacts the number of faults (and thus the resulting logical error rate), it is critical to execute syndrome extraction as quickly as possible.

\textbf{Prior Work and Their Limitations}. Prior work and industry roadmaps \cite{quantinuum2024roadmap, schoenberger2024shuttlingscalabletrappedionquantum, kenpaperSC, 50ionchains} have largely focused on grid-based hardware architectures optimized for topological codes. The connectivity requirements of the quantum error-correcting codes depends on the structure of their stabilizers. For topological codes (like the surface and color code), with local stabilizer checks, a simple 2D grid layout and standard mapping/scheduling method works well~\cite{kenpaperSC}. However, this approach does not generalize to non-topological CSS codes, which often offer higher encoding rate and are thus more favorable for fault-tolerant quantum memory. 
We identify the limitations imposed by previously proposed 2D architectures for executing such codes.
Spatially compact designs with a small number of high-capacity traps\cite{50ionchains} exhibit low levels of parallelism, since typically a single operation can be performed in each trap at a time (as demonstrated in Figure \ref{fig:sensitivity_tight}). On the other hand, sparse grid layouts with many traps (which we designate as the baseline - Figure \ref{fig:background-grids}), are prone to \textit{roadblocking}, where busy traps of one active shuttling path block another, as shown in Figure \ref{fig:fig1}a). These roadblocks significantly reduce realized parallelism, which is particularly problematic for large codes, as the roadblocks in execution results in full serialization of a circuit that is extremely parallelizable. As shown in Figure \ref{fig:noise-effect}, the large amounts of serialization and added depth due to shuttling operations render such codes (Hypergraph product code in this case) unusable on the baseline grid. In contrast, reducing execution time towards ideal parallelism results in orders of magnitude improvements in logical error rates.

\textbf{Improving Parallelism}. To address the roadblocks present in the baseline, we start with a hypothetical, roadblock-free, all-all connected shuttling network and reverse engineer a more practical design from this ideal construction. We find that realizing these designs that fully exploit the potential parallelism is challenging due to constraints on junction and trap connectivity. We categorize roadblocks into costly \textit{trap roadblocks} and less-costly \textit{junction roadblocks}, and find a \textit{junction network} design that eliminates all trap roadblocks in favor of junction roadblocks. However, junction roadblocks remain sub-optimal due to high spatial overheads associated with junction crossing. Inspired by the mesh network, we introduce a simple yet effective architecture that eliminates all roadblocks, achieves high parallelism, and keeps spatial and control overhead low. 

\textbf{Cyclone: Achieving Roadblock Free Parallelism with Low Spatial and Control Overhead} We propose a completely roadblock-free, highly parallel codesign that bounds the number of time steps to be linear in the number of stabilizers in a code, we designate as \textit{Cyclone}. The Cyclone codesign consists of two distinguishing features:
\begin{enumerate}
    \item \textit{Hardware} must support a topological ring-like connectivity with at most $\frac{m}{2}$ traps, where $m$ is the number of stabilizers
    \item \textit{Software} must have symmetric and parallelized shuttling operations
\end{enumerate}

Cyclone fixes a direction and symmetrically moves all ancilla qubits in lockstep across balanced partitions of data. Once two full rotations around the loop are complete, the code finishes execution. Due to its simplicity, Cyclone achieves lower spatial overhead as compared to the baseline, minimal control overhead (theoretically requiring only one DAC with forwarding --- in practice, wiring complexities across large machines could make this number a bit higher), and highly parallelizable structure. Notably, the baseline requires only slight modifications to satisfy the hardware requirement, as shown in Figure \ref{fig:cyclone_base_design}b, demonstrating the flexibility of this codesign. We refer to the obvious circular topology with L-shaped junctions (Figure \ref{fig:cyclone_base_design}a) as the \textit{Base Cyclone} codesign, with exactly $\frac{m}{2}$ traps. Depending on the specific values of the shuttling and gate times, as well as the code's stabilizers, denser or more optimized versions of Cyclone may be achievable (Figure \ref{fig:sensitivity_tight}). The base Cyclone codesign beats the baseline grid $n \times n$ by up to $2-3$ orders of magnitude in logical error rate, and achieves approximates $\sim 20 \times$ better spatial efficiency combined with order $O(n^2)$ simpler control overhead.

Our contributions are as follows: \begin{enumerate}
    \item We identify that conventional 2D-grid codesigns are insufficient for non-topological CSS codes that require non-uniform and long-range interactions, often creating high amounts of roadblocks and thus exhibiting low levels of parallelism.
    \item We find other codesigns that capture this parallelism much better and conclude that they come with design constraints restricting their practical feasibility. 
    \item We propose an efficient software-hardware codesign designated as Cyclone that bounds the overall syndrome extraction latency, achieves high level of parallelism, has great spatial efficiency, and minimal control overhead.
\end{enumerate}

\begin{figure*}[htbp!]
    \centering
    \includegraphics[width=0.95\linewidth]{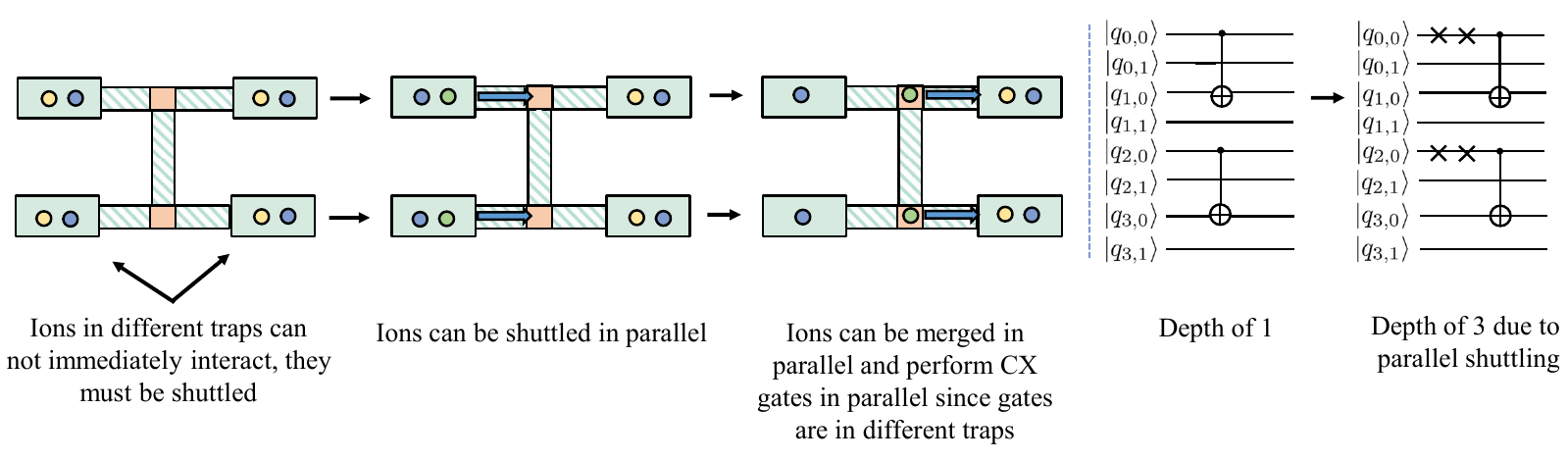}
    \caption{From left to right and top to bottom, qubit labelling follows the ordering (trap number, ion number). The yellow qubits are to interact in this piece of the circuit. The ancilla bits (now moving and highlighted in green) are shuttled towards a junction in one step, across the junction, and then are merged into the target data trap. This process, along with the ensuing gates, is performed in parallel, and the new circuit is shown (X denotes a shuttling operation).}
    \label{fig:QCCDShuttling}
\end{figure*}

\section{Background}

\subsection{QEC Codes}
Quantum error correction (QEC) codes~\cite{fowler2012surface, kitaev1997toriccodes, gottesman1997stabilizer, landahl2011colorcodes} protect quantum information from noise by encoding it across multiple physical qubits and using noisy syndrome outcomes to detect and correct errors.
Here, we focus on \textit{CSS stabilizer codes},~\cite{fowler2012surface, Grassl_2023, landahl2011colorcodes, ibmBB} a widely used family of QEC codes defined by a set of Pauli operators known as stabilizers.
In particular, we study two families of CSS stabilizer codes: hypergraph product codes~\cite{HGP_discovery, quitsAlgorithm, quits} and bivariate bicycle codes~\cite{ibmBB, BB_code_logical,BBcodes_more}.
The former represents an \textit{edge-colorable}~\cite{quits, quitsAlgorithm} code, enabling interleaved measurement of $X$ and $Z$ stabilizers, while the latter is \textit{non-edge-colorable} and does not allow such interleaving.
In what follows, we describe the syndrome extraction process and provide details on the specific codes used in our study.

\subsubsection{Syndrome Extraction Circuit}
The syndrome extraction circuit measures a set of stabilizers (Pauli operators) derived from the code’s parity check matrix. 
These circuits involve entangling data qubits with ancilla qubits, followed by measurements of the ancilla qubits to infer the presence and type of errors affecting the encoded quantum state.
A classical decoder then processes the syndrome outcomes to determine a correction that restores the logical state.

However, the syndrome extraction circuit is itself faulty as it is subject to both \textit{operation errors} (e.g., gate and measurement errors) and \textit{latency-induced errors} (e.g., decoherence during idle periods).
Consequently, the classical decoder uses noisy syndrome data to find the correction, and depending on the code, it may require one or multiple rounds of measurements to determine the correction.
The overall \textit{logical error rate} is therefore affected by both the fidelity of the syndrome extraction circuit and the effectiveness of the decoder.
\textbf{In this paper, we focus on improving the syndrome extraction circuit by reducing latency, as the gate and measurement operations are identical across compilers and architectures.}

\subsubsection{Hypergraph Product Codes}
Hypergraph Product (HGP) codes~\cite{HGP_discovery, quits, quitsAlgorithm} are a class of CSS stabilizer codes constructed by taking the product of two classical codes. 
They belong to the broader family of quantum low-density parity-check (qLDPC) codes and are known for their favorable properties, such as low-weight stabilizers, asymptotically good distance, and potentially high encoding rates. 
Importantly, HGP codes are \textit{edge-colorable},~\cite{quits, quitsAlgorithm} allowing $X$ and $Z$ stabilizers to be interleaved in the syndrome extraction schedule, enabling highly parallel implementations that reduce circuit latency. 
However, syndrome extraction in HGP codes requires long-range interactions, making them challenging to implement on standard 2D grid-based architectures, making them a good target for this paper. 

\subsubsection{Bivariate Bicycle Codes}
Bivariate Bicycle (BB) codes~\cite{ibmBB, BB_code_logical,BBcodes_more} are a class of CSS stabilizer codes constructed from bivariate polynomials A and B.
These codes have several desirable properties similar to other qLDPC codes, including relatively low-weight stabilizers, and high encoding rates and good distance.
Unlike HGP codes, BB codes are not \textit{edge-colorable}, which prevents interleaving of $X$ and $Z$ stabilizer measurements and slightly limits the parallelism achievable during syndrome extraction.
However, like HGP codes, BB codes require long-range interactions and are not well-suited for implementation on naive 2D grid-based architectures.
These characteristics make BB codes a good and contrasting choice to study in our design space exploration.

\subsubsection{Motivational Case Study: Architectural Need for High Parallelism}
\begin{figure}[htbp!]
    \centering
    \includegraphics[width=0.9\linewidth]{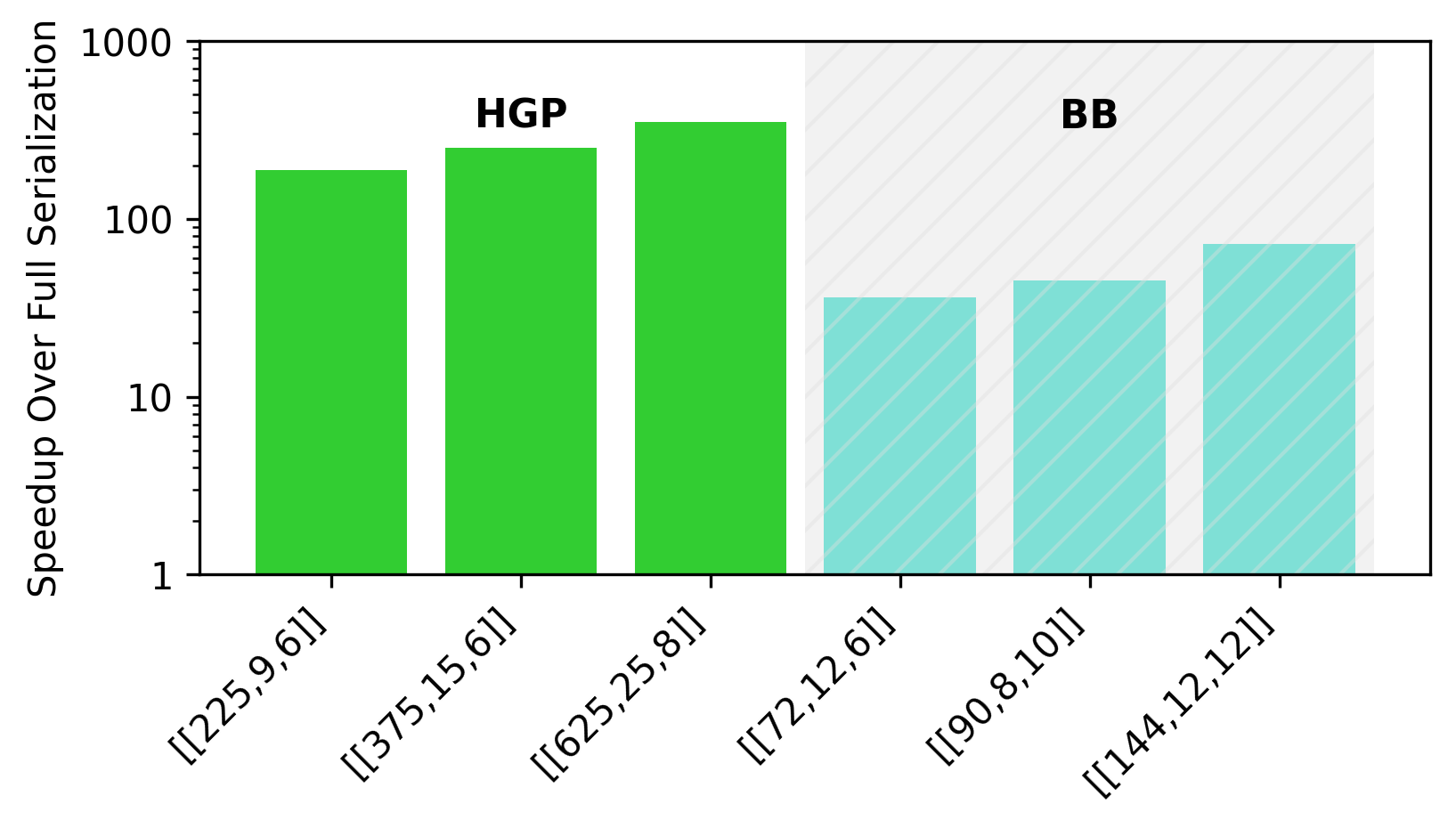}
    \caption{Comparing Speedup of fully parallel vs. fully serial designs for the two classes of codes considered in this paper. Bars show a relative speedup ($\times$ times) over a fully serialized version of the code.}
    \label{fig:comparing_parallelism}
\end{figure}

Both HGP and BB codes are highly parallelizable. In Figure \ref{fig:comparing_parallelism}, we compare the latency of the maximally parallel schedule (constructed using the method described in Section \ref{sec:finding-ideal-parallel}) with that of a fully serial schedule, and observe a dramatic speedup. In addition, this speedup scales asymptotically well with increasing code size, indicating that executing fault-tolerant circuits with minimal latency, and thus minimal logical error, requires QCCD architecture that can support large amounts of parallelism. \textbf{For topological codes such as the Surface and Color Codes, the gridlike QCCD structure is already fast and sufficient} \cite{landahl2011colorcodes, paulitwirl2, lee2024concatmwpm}. However, for non-topological CSS codes with potentially better encoding rates and thresholds, like the ones considered in this work, identifying optimal QCCD architecture remains an open problem.

\subsection{QCCD Systems}

QCCD Trapped-ion systems offer a wide range of tunable parameters and architectural design choices, which we explore through simulation. We fix shuttling and gate overheads and then compile a circuit to a target architecture, where \textit{inter-trap} parallelism is unconstrained, while \textit{intra-trap} parallelism is limited to one gate per trap at any given time.

\subsubsection{Shuttling Overhead}
\label{sec:shuttling-overhead}

Shuttling operations in QCCD systems are composed of atomic steps: \textit{split}, \textit{move}, and \textit{merge}. These operations allow ions to depart from one trap (split, 80 $\mu$s), move across shuttling zones (10 $\mu$s), and combine with another trap (merge, 80 $\mu$s). All shuttling operations can be parallelized across the entire system. Since, the relative order of ions within a chain typically needs to be changed one or more times throughout a shuttling operation, so SWAP operations are inserted to switch their relative positions. SWAP operations can be one of two types: either position-based swaps within a chain (2–4 ion SWAPs, constrained by ion proximity), or \textit{GateSwaps} implemented using three CX gates, allowing swaps between arbitrary ion positions. In this work, we consider GateSwap for our evaluation, as the overhead is dependent on fixed gate times as opposed to arbitrary interaction distance in a chain length. We study sensitivity to this choice later in Section \ref{sec:sensitivity}. 

Shuttling is unidirectional, and each trap connects to at most two shuttling paths. To increase connectivity, junctions (buffers where ions can easily be transported across) are used. Junctions can have a degree of connectivity up to 4, and the crossing time is dependent on the degree, 10, 100, and 120 $\mu$s for degrees 2, 3, and 4, respectively~\cite{murali} . If a trap exceeds its ion capacity during a given shuttling operation, a costly \textit{rebalance} is triggered. Rebalances aim to reduce congestion in nearly full traps such that shuttling operations can resume without exceeding the trap capacity. These are inserted into the schedule on an ``as-needed" basis throughout compilation, and take nontrivial amounts of time to perform.

\subsubsection{Compilation}
A QCCD compiler maps program qubits onto machine qubits, routes them across the physical architecture, and schedules both gates \textit{and} shuttling operations according to hardware constraints, all with the goal of minimizing the overall circuit latency. A sample schedule is shown in Figure \ref{fig:QCCDShuttling}, where ions are shuttled from traps 0 to 1 and traps 2 to 3, crossing a junction and then undergo a CX gate between the respective data and ancilla pair. QCCD compilers use greedy heuristics \cite{murali, Saki_2022} such as the Earliest Job First (EJF) policy in order to minimize the shuttling overhead, but typically generate uncoordinated schedules and limited parallelism.

Compilation is focused on optimizing overall circuit latency, which is especially important in the case of optimizing QEC syndrome extraction circuits, as latency has a direct impact on injected decoherence error. Larger decoherence error creates faultier syndrome extraction circuits, which contribute to overall increased logical error.

\subsubsection{Topologies}

\begin{figure*}[htbp!]
    \centering
    \includegraphics[width=0.9\linewidth]{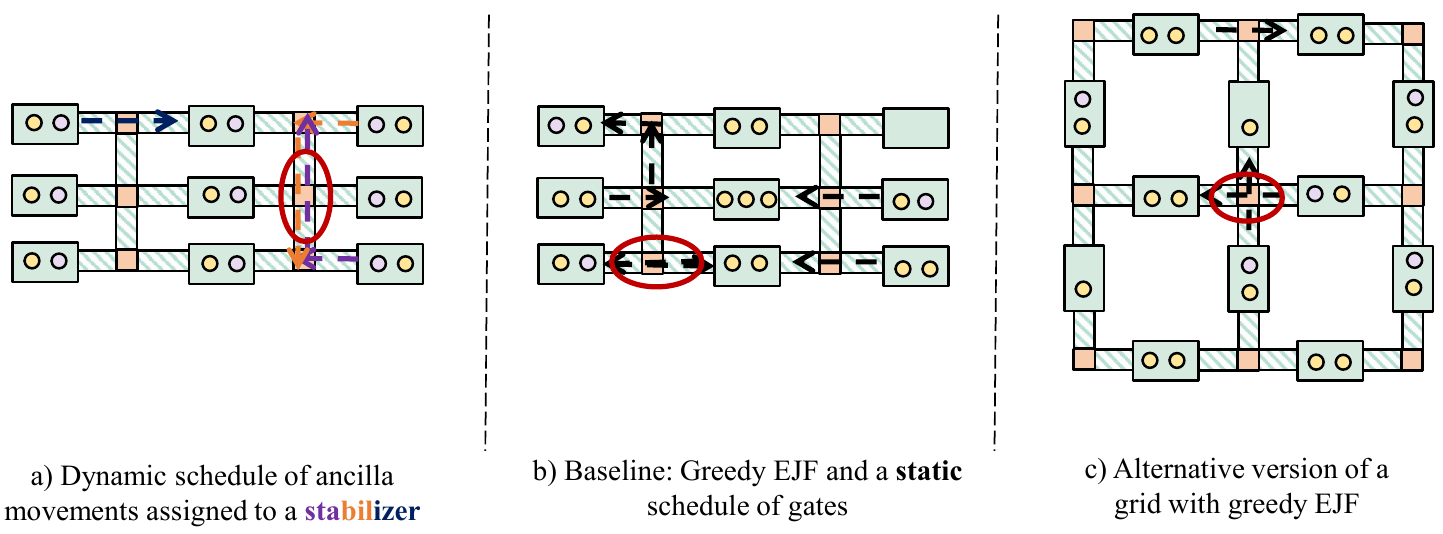}
    \caption{We compare the 3 different grid inspired codesigns used in this paper a) The dynamic schedule on a grid, which we find performs worse than the baseline due to heavy roadblocking b) a static EJF schedule with a greedy cluster mapping on a grid, the baseline \cite{murali} c) an alternative version of a grid with alternating horizontal/vertical meshes and L-shaped junctions.}
    \label{fig:background-grids}
\end{figure*}

Trapped-ion device topologies are constrained to a single physical plane, leading many designs to adopt mesh-like grid structures. However, it is important to emphasize that \textbf{not all grids are created equal.} The fundamental differences between junctions (which allow for connectivity up to degree 4) and traps (which allow connectivity up to degree 2), along with their interleaving arrangements give rise to different types of grids with distinct properties. In Figure \ref{fig:background-grids}, we show the standard grid (b) that allows for flexibility in the vertical direction, and an alternate version of a grid (c) with L-shaped junctions allowing for easy circular paths, which allows the surface code to be performed trivially \cite{kenpaperSC}. These are both considered grids, but each come with their own tradeoffs. In Figure~\ref{fig:background-grids}a), we show how different software policies yield different codesigns and thus different results. Notably, dynamic scheduling of ancilla qubits assigned to stabilizers (as described in Section \ref{sec:finding-ideal-parallel}) when paired with a grid codesign suffer from excessive roadblocking, leading to \textit{worse} performance than the greedy software technique of mapping and scheduling based on clusters of local interactions. Based on this insight, we use the static scheduling approach in \cite{murali} combined with the hardware design more flexible on vertical directions as \textbf{the baseline for the rest of the paper}. Our baseline, inspired by industrial designs, is a generalized version of the grid implemented in \cite{oxfordionicsgrid}, augmented with additional columns of vertical junctions between each trap to enhance its capability to combat roadblocks. The alternate grid design follows the same architecture as presented in \cite{Dac-forwarding}, scaled to an arbitrary size. Both serve as representative roadmaps for grid designs commonly employed in industry. 

\subsubsection{Wiring QCCDs}

QCCDs typically require one Digital-to-Analog Converter (DAC) per trap to generate the analog control signals necessary to control and move ions. If identical ion movements occur across multiple traps, the same control signal can be cowired or \textit{broadcasted} to multiple traps~\cite{honeywell_racetrack, Dac-forwarding, osti_1237003}. This approach eliminates the need for duplicating DACs and their associated wiring, significantly simplifying the qubit control infrastructure and reducing hardware overhead~\cite{Dac-forwarding, DAC_savings}.

\subsection{Realistic Noise Models for Hardware Simulation}
\label{sec:noise-model}

We aim to construct hardware-aware noise models for our simulation of logical error rate. We do this by augmenting the conventional circuit level noise model with a depolarizing channel, to model decoherence due to circuit latency using the Pauli twirling approximation \cite{paulitwirl1, paulitwirl2}.
\subsubsection{Base Noise Model}
The base noise model consists of multiple error sources, including single qubit and two qubit gate errors, state preparation errors, and measurement errors.
These are modeled using stochastic depolarizing channels, with each error occurring independently with probability p, also referred to as the physical error rate.

\subsubsection{Modeling Decoherence: Latency Couples with Error}
To account for time-based decoherence, we use the Pauli twirling approximation \cite{paulitwirl1, paulitwirl2} to model time based error as a depolarizing channel using both decay time (T1) and dephasing time (T2). We assume coherence times can be anywhere in the range of 10-100 seconds, consistent with present-day trapped-ion devices. We parameterize this coherence time based on \textit{p} with a log fit from $p = 10^{-4}$ (100s) to $p = 10^{-3}$ (10s). This parameterized coherence time is then used for both $T_1$ and $T_2$, denoted as $T_a$ and $T_b$, respectively. Together with the circuit execution latency $t_E$, these values are used in the Pauli twirling approximation \cite{paulitwirl1, paulitwirl2} to derive an effective depolarizing noise model for decoherence.

\begin{figure}[htbp!]
    \centering
    \includegraphics[width=0.9\linewidth]{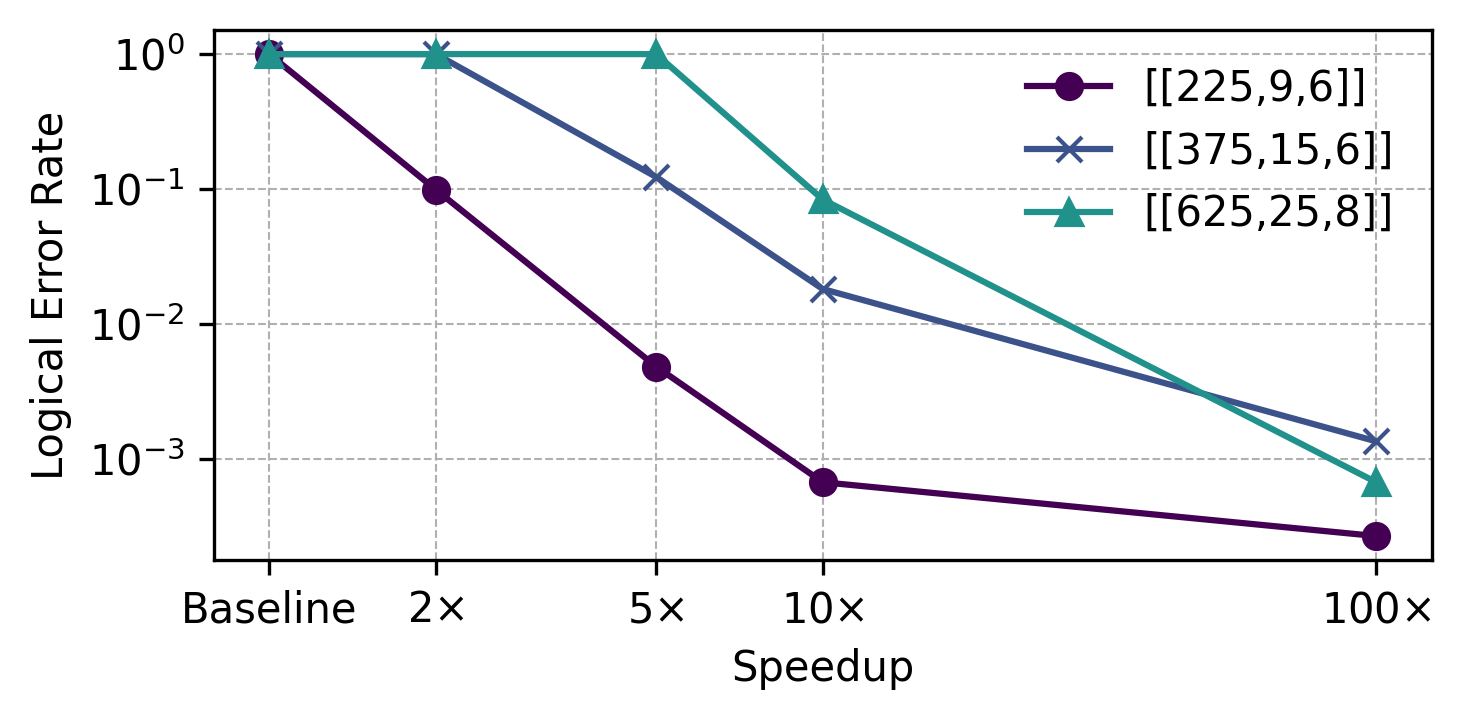}
    \caption{Comparing the logical error rate improvement upon speeding up the Baseline across different HGP codes. Physical error rate is held constant at $p = 5 \times 10^{-4}$. For all of these HGP codes, the baseline puts the code above the threshold.}
    \label{fig:noise-effect}
\end{figure}

We then combine the base noise model with the Pauli twirling approximation noise to create a hardware-aware noise model. As shown in Figure \ref{fig:noise-effect}, we find that just a 2$\times$ reduction of depth yields a lower logical error rate by $\sim90\%$. 

\subsection{Limitations of Prior Work and Need for Codesign}

\begin{figure}[htbp!]
    \centering
    \includegraphics[width=0.9\linewidth]{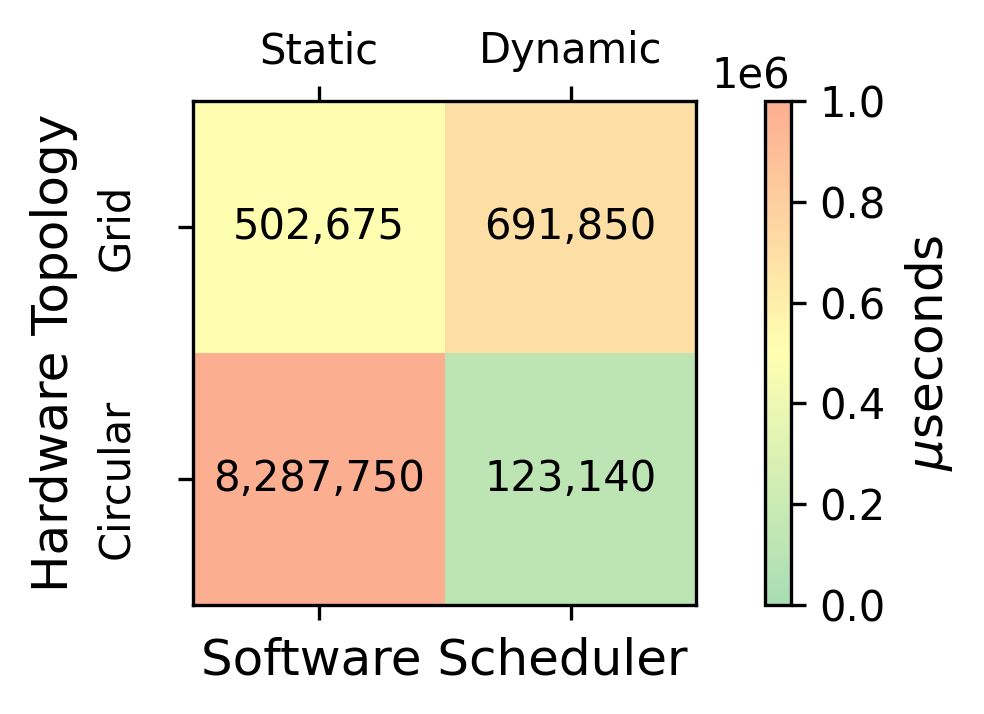}
    \caption{Confusion Matrix for software type (dynamic or static, depending on whether or not schedules are interpreted as a set of parallelizable gate timeslices as described in Section \ref{sec:finding-ideal-parallel} or a DAG from an input reader), or connection topology being a circle or a grid.}
    \label{fig:heatmap_codesign}
\end{figure}

The current grid-based hardware design imposes severe limitations on performance. As shown in Figure \ref{fig:heatmap_codesign}, the confusion matrix highlights that changing the hardware or software alone is insufficient for realizing the maximum parallelism and minimizing logical error rates. The grid-based hardware with an optimized dynamic scheduling software policy leads to a very large hardware execution time due to roadblocks, even higher than the baseline using a greedy EJF schedule. Similarly, a sparse circular hardware topology with a greedy EJF static schedule is disastrous. A \textit{coordinated}, roadblock-free software policy (which allows gates to be scheduled dynamically) along with a hardware topology designed to mitigate roadblocks can solve these problems. \textbf{Optimizing both hardware and software to create codesigns is a necessary step for efficient fault tolerant quantum memory}.    

\section{Challenges In Building Parallel Designs}
In this section, we explore the different methods of achieving practical QCCD designs that aim to exploit the high levels of parallelism in Hypergraph Product Codes and Bivariate Bicycle codes. We begin with an idealized design, referred to as OPT, which assumes maximal parallel design but is spatially impractical due to the dense shuttling connectivity. From this theoretical baseline, we reverse engineer practical solutions given design constraints. We outline two types of designs that attempt to retain some of the efficiency in OPT, but are still impractical due to current junction design constraints. Finally, we use the insights gained from analyzing these tradeoffs to motivate Cyclone, introduced in Section \ref{sec:cyclone}.

\subsection{Finding The Maximum Bound for Parallelism}
\label{sec:finding-ideal-parallel}

To determine the maximum achievable parallelism, independent of hardware constraints, we analyze the structure of the QEC circuit. We use the two distinguishing classes (in their respective order of granularity) that could create tighter conditions and therefore more maximally parallel conditions: \textit{Edge Colorable CSS Codes} and \textit{Non-Edge Colorable CSS Codes}. In this work, we use Hypergraph Product Codes as a representative of the former, and Bivariate Bicycle Codes as a representative of the latter. 

\begin{enumerate}
    \item \textit{Edge-colorable QLDPC codes} This category includes codes such as Hypergraph product codes, Quasi-cyclic lifted product codes, Balanced product codes, all of which are created from the homological product of two classical codes. By construction, the edges of the connectivity graph can be split into cardinal directions, allowing for interleaved $X$ and $Z$ stabilizer measurements that produce a tighter bound on the maximal parallelism \cite{quitsAlgorithm}. All of these are Calderbank-Shor-Steane (CSS) codes. Most QLDPC codes that do not fall into this category typically belong to the next group of more general CSS codes.
    \item \textit{Non Edge-colorable CSS Codes}. In any CSS Code, each individual stabilizer is composed entirely of either $X$ or $Z$ Pauli operators, never a mix. This allows for a straightforward parallelization strategy: all $X$ stabilizers can be executed in parallel, followed by all $Z$ stabilizers (or vice versa). This leads to a schedule that has a worst case depth of $w_{max}(m_x) + w_{max}(m_z)$ where $w_{max}(m_x)$ and $w_{max}(m_z)$ denote the maximum weight of any $X$ and $Z$ stabilizer, respectively.
\end{enumerate}

We can use the corresponding policies to generate timeslices, where our maximally parallel compiler iterates through each timeslice and schedules all gates freely within each individual timeslice, therefore producing a \textbf{``dynamic"} schedule as opposed to a fixed DAG representation. 

\subsection{Constructing Maximally Parallel Hardware}

\begin{figure*}[htbp!]
    \centering
    \includegraphics[width=0.9\linewidth]{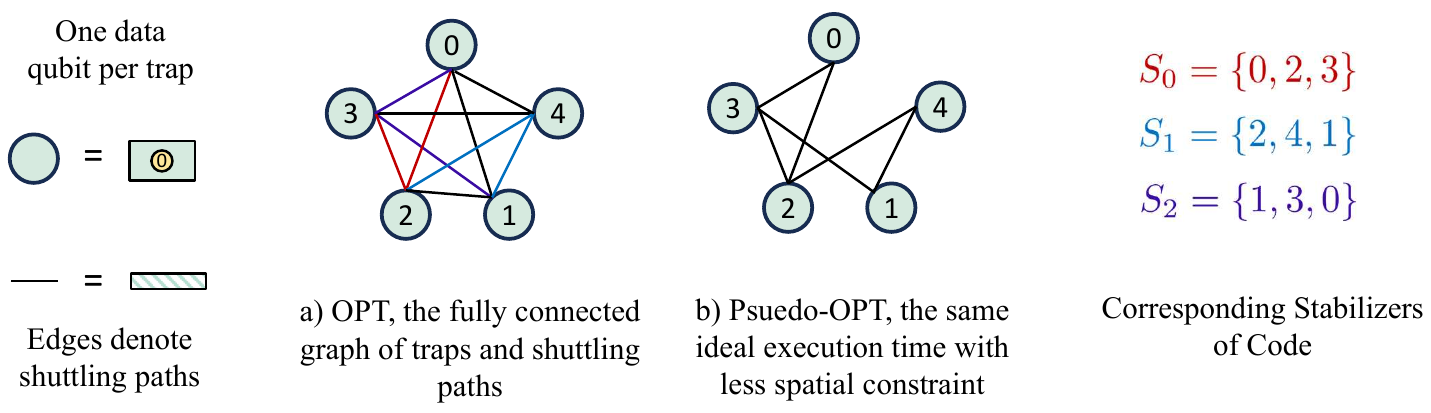}
    \caption{Traps are represented as nodes containing one data and edges represent shuttling connection paths. OPT and Psuedo-OPT are shown in a) and b) executing the stabilizer schedule shown on the right. The corresponding edges used are highlighted in OPT, and the ones not used in OPT are pruned in Psuedo-OPT.}
    \label{fig:grip_stage1}
\end{figure*}

To construct a maximally parallel design, we begin with an idealized configuration (designated as OPT), in which the system is modelled as a fully connected graph where the nodes represent traps, and the edges represent shuttling paths. In this architecture, each data qubit is assigned a dedicated trap, allowing ancilla qubits to move between them with minimal interference, thereby achieving the lowest possible execution latency as determined by the ideal parallel times discussed in Section \ref{sec:finding-ideal-parallel}. Because this configuration has effective all-to-all connectivity, it avoides shuttling congestion, and completes syndrome extraction in the shortest amount of timesteps possible, barring two caveats:
\begin{enumerate}
    \item Traps can have arbitrarily high degree
    \item Multiple edges connecting the same set of nodes can only be simplified to one edge if there is no bidirectional traffic.
\end{enumerate}
The first of these problems can be solved by adding more junctions as postprocessing. The second problem can be solved because the usage of the same edge at the same time is extremely rare, and if it were to happen it would cause a small stall as the move time between traps is much smaller than the split/merge, and gate times, $m_t << s, g$. However, due to its non-planar configuration, as shown in Figure \ref{fig:grip_stage1}, OPT is almost always not possible, and its structure must be altered.

We first prune OPT to create pseudo-OPT (Figure \ref{fig:grip_stage1} part b), where all unused edges in OPT are discarded (i.e. retaining only those shuttling paths necessary to connect data qubits involved in a given stabilizer measurement). While Pseudo-OPT preserves the ideal runtime, and improves spatially efficiency, it remains highly non-planar and is not physically realizable for any of the codes studied in this paper. 

\subsection{Mesh Junction Network}
\begin{figure}[htbp!]
    \centering
    \includegraphics[width=0.5\linewidth]{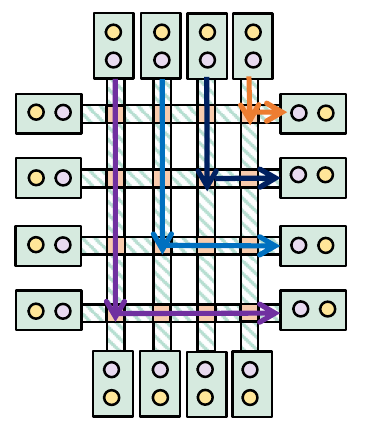}
    \caption{Network of junctions allowing up to $\frac{n}{4}$ synchronous interactions, where the value of $n$ is fixed to 16 in the figure.}
    \label{fig:mesh-junction-network}
\end{figure}

Eliminating edges or merging nodes in Pseudo-OPT introduces roadblocks similar to that in the baseline grid architecture. However, adding junctions as ``free" nodes with a maximum degree of 4 to achieve all to all connectivity is possible. In Figure \ref{fig:mesh-junction-network}, we show how adding a dense network of junctions in between all the traps can help establish this all-to-all connectivity. We examine a perfect junction mesh of size $\frac{n}{4} \times \frac{n}{4}$, where $n$ is the number of traps in Pseudo-OPT (equal to the number of data qubits). With traps laid out in this way, the worse case time complexity is $2 \times \frac{n}{4} = \frac{n}{2}$ junctions, as this is equivalent to the Manhattan distance on a square grid. This design converts all of our \textit{trap-roadblocks}, which require expensive SWAP gates, into \textit{junction-roadblocks}, which are easier to handle. However, junction-roadblocks and crossing times (of degree 4 junctions) are still nontrivial, especially at scale. This leaves two problems with the dense network of junctions design:

\begin{enumerate}
    \item \textit{Temporal Constraints: Path Collisions at Junctions}.  This design lacks a simple way to resolve junction-junction collisions. As a result, the compiler must use conservative path scheduling. In an optimistic case where $\frac{n}{4}$ paths are scheduled at a time (i.e.  $\frac{n}{4}$ gates executed in parallel, and they are not all short paths) as shown in Figure \ref{fig:mesh-junction-network}, there is still $\prod_{m_i}^m w(m_i)$ gates to execute. Junction crossings are cheaper than shuttling operations and gate operations, but as $\frac{n}{2} - 1$ high degree junctions are hit per time slice, the cost $(\frac{n}{2} -1) \times j_c$ (junction crossing time) is significant. If high-degree junction crossing times become as cheap as movement costs, the temporal cost may not be much of an issue, and a mesh-like junction network may be a good choice if temporal optimization is a priority. 

    \item \textit{Spatial Constraints: Junction Overhead} The number of junctions scales quadratically with the code size, $(\frac{n}{4})^2$, posing real scalability issues. For some of the codes considered in this paper this translates to over several thousands of junctions, which is already impractical. 
\end{enumerate}

\begin{figure}[htbp!]
    \centering
    \includegraphics[width=0.9\linewidth]{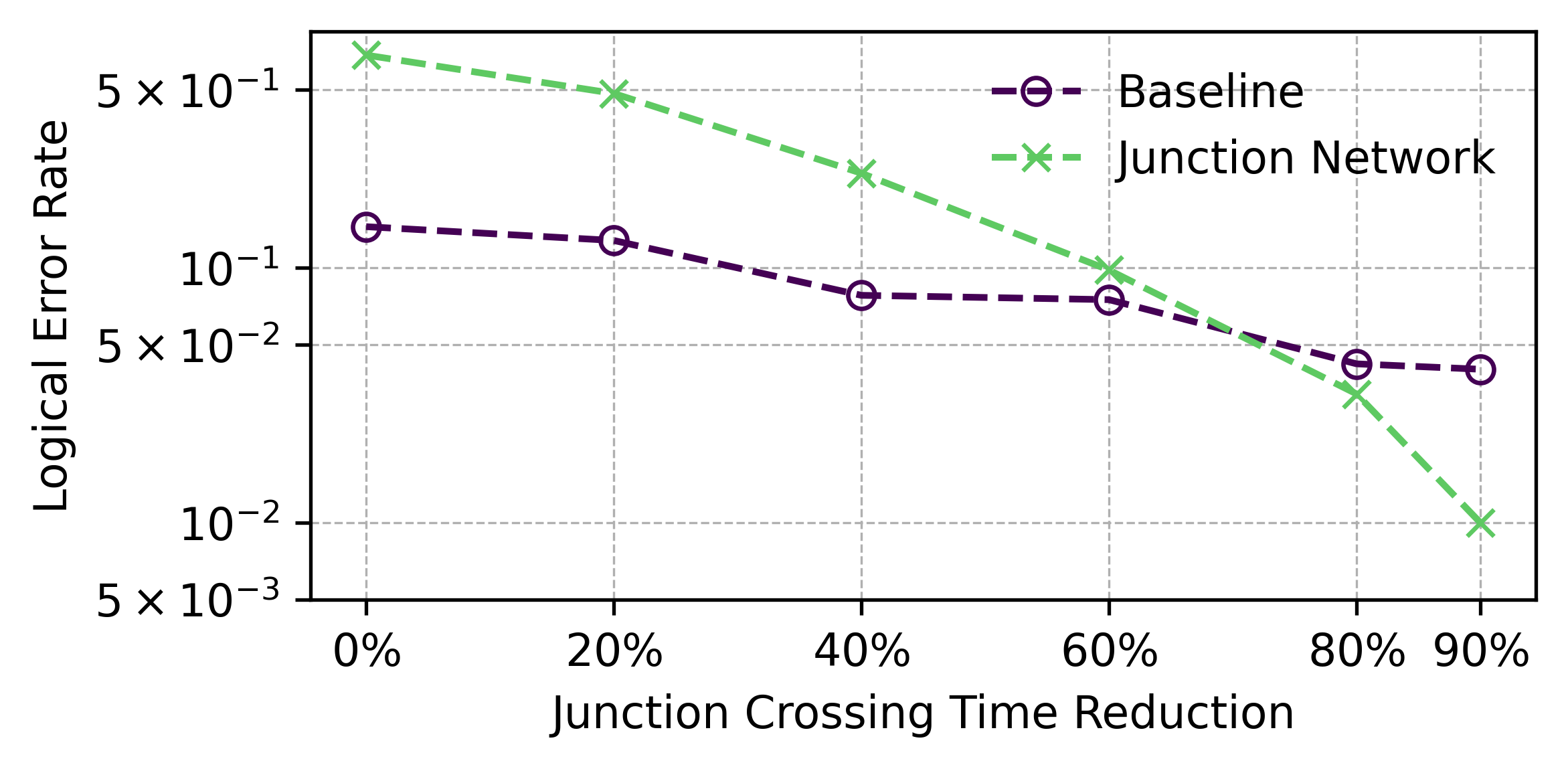}
    \caption{Logical Error Rate across levels of optimistic assumptions in junction crossing times. At around $70\%$ reduction, the junction network design becomes temporally feasible.}
    \label{fig:junction-network-plot}
\end{figure}

\begin{figure*}[htbp!]
    \centering
    \includegraphics[width=0.9\linewidth]{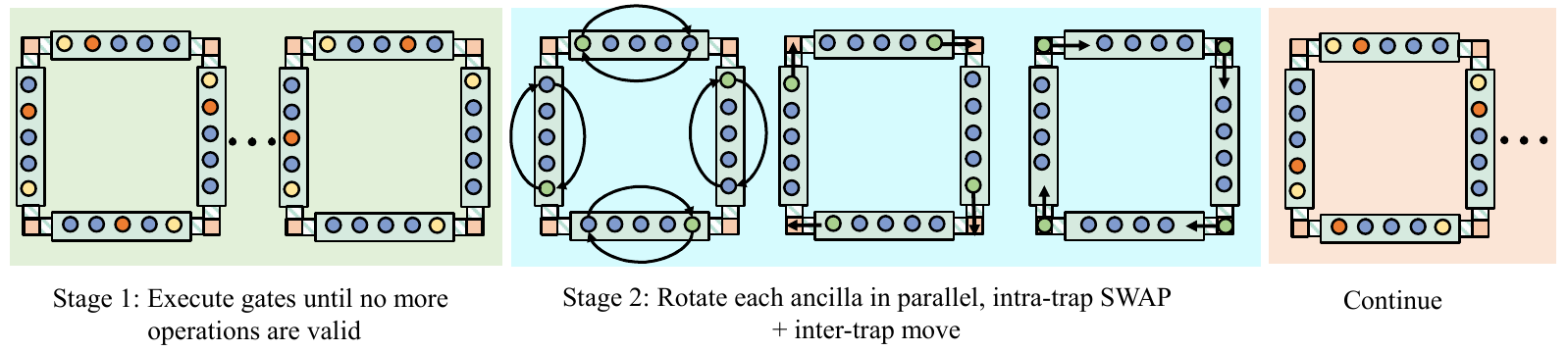}
    \caption{The three stages of the Cyclone design: The initial start state of Cyclone. Ancilla are marked in yellow, and active gates upon data qubits are marked in orange. All gates across the machine at a single time step are parallelized. In the above example, there are up to two qualifying data qubits satisfying the conditions of being in the respective ancilla's parity check and being in the same trap, so this whole stage takes the total time of 2 gates. In stage 2, the ancilla (now highlighted in green) are swapped with a data qubit (blue) at the traveling edge of the trap. These can all be parallelized by enforcing gate swaps (3 CXs). The ancilla are then all simultaneously split into a shuttling zone and the ancilla are then merged into a new trap. The design then continues round robin until all ancilla have traveled to all traps, identical to the starting configuration.}
    \label{fig:cyclone_process}
\end{figure*}
If junction crossings were significantly faster, i.e., $j_c << g$, then even though movement through the network must occur serially, the speed of the movement compared to the gate time would allow a future gate to be scheduled (which has overlapping junction crossings to the ones used in the previous gate) while the previous gate is still in execution. This is effectively more parallel gate execution, as gate times overlap more. In Figure \ref{fig:junction-network-plot}, we analyze the sensitivity of performance to junction crossing $j_c$ and find that a $70\%$ reduction in junction crossing times would allow the mesh network to outperform the baseline grid. However, while this would be temporally optimal, it still suffers from quadratic scaling in terms of the spatial requirement of junctions, which is not ideal for large codes like the $[[625,25,8]]$ Hypergraph product code.

\section{Cyclone: Towards Spatially Efficient and Practical Parallel Architectures}
\label{sec:cyclone}

\begin{figure}[htbp!]
    \centering
    \includegraphics[width=0.9\linewidth]{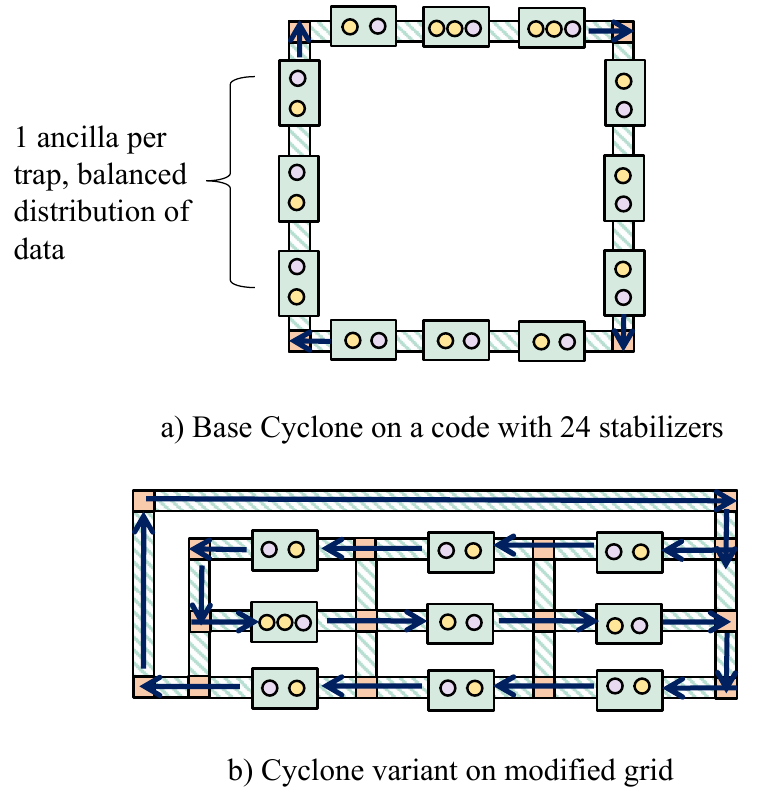}
    
    \caption{a) Base cyclone for a code with ancilla distributed 1 per trap, and data distributed in the most balanced method (if $\frac{m}{2} \mid n$, all traps will have equal amount of data). b) Implementing Cyclone by slightly modifying a grid, although due to extra junctions in the last connection, ions in remaining parts of the machine must stall on each step and wait for the ion to pass through the long connection to preserve symmetry.}
    \label{fig:cyclone_base_design}
\end{figure}

We can reduce most of the junction overhead from the mesh-junction network by retaining only the outer trap ring and enforcing a fixed ancilla movement orientation. In this simplified architecture, only L-type junctions are needed, and are placed on the corners of the circle. Ancilla qubits then move in a round robin fashion around the circle, performing a gate only if one or more of the data qubits it is touching is part of its assigned stabilizer. Stabilizers are dynamically assigned to ancilla as per the \textit{non-edge colorable} dynamic schedule paradigm (\textit{regardless} of HGP or BB code), all $X$ stabilizers are measured during the first complete rotation, followed by all $Z$ stabilizers during the second. We designate this circular hardware layout, together with this precise and symmetric software policy as the \textit{Cyclone} codesign. Cyclone guarantees zero roadblocks, bounds total movement, and maintains high parallelism. The number of traps in the \textit{base form of Cyclone} is set to the $max(|\mathbf{X}|,|\mathbf{Z}|)$, where $\mathbf{X}$ and $\mathbf{Z}$ represent the set of $X$ and $Z$ stabilizers, respectively. Typically, the size of these sets is equal, and we divide the set of stabilizers of size, $m$, into two sets of equal size ($\frac{m}{2}$) (for the codes in this paper, this is true, and we will use this value for further analysis in runtime). The symmetric property guarantees that upon two full rotations around the circle (one for $X$ stabilizers and one for $Z$ stabilizers), a single round of syndrome extraction must be complete. A walk-through of a single time step in Cyclone is shown in Figure \ref{fig:cyclone_process}. In Figure \ref{fig:cyclone_base_design}a), we show the base form of Cyclone for a code with twelve $X$, and twelve $Z$ stabilizers. Since the ancilla are reused between $X$ and $Z$ stabilizer measurement rounds, only $\frac{m}{2}$ ancilla qubits are needed, reducing both the number of traps and qubits. In this section, we detail how this codesign yields faster execution, spatial efficiency, easier control overhead, and better logical error rate as compared to the baseline.

\subsection{Runtime and Spatial Efficiency Analysis}

The primary objective of the Cyclone architecture is to exploit the high levels of parallelism of HGP and BB codes to achieve fast execution time (roughly $O(m)$ timesteps) while still having a practical, spatially efficient design ($O(m)$ in terms of traps and connections).

\begin{figure}[htbp!]
    \centering
    \includegraphics[width=0.95\linewidth]{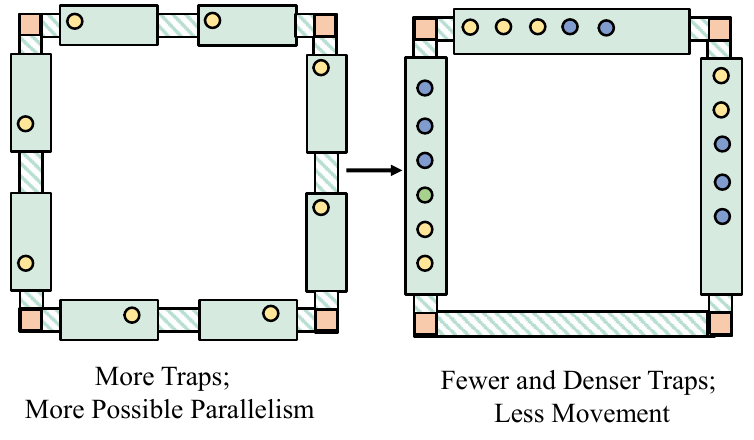}
    \caption{Compressing Cyclone to a more spatially efficient and practical design. In this example, we have purposefully shown the case of an asymmetry, taking a base design of 8 traps into 3. In this rotation, an ancilla still must carry out its third gate (green) and all other ancilla qubits (yellow) must wait until this operation completes to proceed and rotate to the next trap. All ancilla still move across the traps in unison. Data qubits are shown in blue.}
    \label{fig:cyclone_trapfolding)_diagram}
\end{figure}

\textbf{Cyclone is flexible in the order of traps/connections}, we can extend the base cyclone construction to fit on any arbitrary cycle graph $C_x$ where $x \neq \frac{m}{2}$. In this flexible model, $\frac{m}{2}$ ancilla must be distributed evenly (or as even as possible) across all the $x$ traps. The worst case gate time per each rotation of Cyclone becomes $\lceil\frac{m}{x}\rceil$, reducing parallelism as $x$ decreases. If $x \nmid m$, imbalances arise and some traps will hold more ancilla than others, and each gate step of the Cyclone process may have to wait for the trap with the most ancilla to finish as shown in Figure \ref{fig:cyclone_trapfolding)_diagram}. This temporal inefficiency is typically not too costly, because data partitions are already somewhat uneven in practice. Even the base version of Cyclone may be performing between $0-\frac{2n}{m}$ gates on each trap in a single time step, causing some stalling. Keeping balanced partitions and an amount of traps close to $\frac{m}{2}$ makes this disparity sufficiently low, as for all codes in this paper, $m \sim n$. Finally, we opt for a non-edge-colorable dynamic schedule rather than the edge-colorable schedule because the non-edge-colorable method guarantees circuit completion in exactly 2 full rotations of the Cyclone loop, whereas the edge colorable method requires more, between $8-12$ for the HGP codes in this paper.

\begin{figure}[htbp!]
    \centering
    \includegraphics[width=0.95\linewidth]{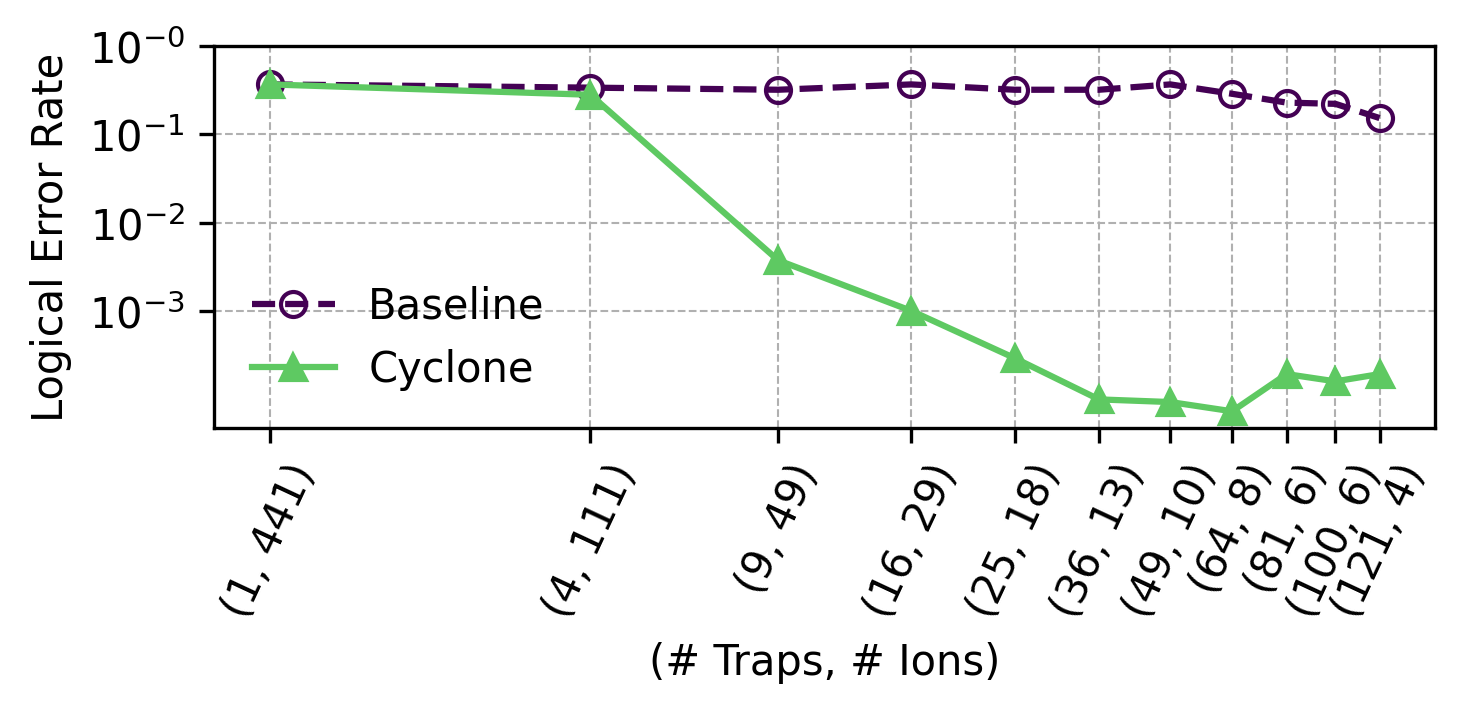}
    \caption{Comparing Sensitivity to different Trap/Ion arrangements at a constant physical error rate of $10^{-4}$ on a $[[225,9,6]]$ code.}
    \label{fig:sensitivity_tight}
\end{figure}

Condensing Cyclone from its base form can offer notable advantages. Since trap count and ion capacity are tightly coupled, the total number of qubits in the system being the product of the two, adjusting this balance impacts both spatial and temporal efficiency. In our previous experiments, we assumed baseline ion trap capacity to be a constant, roughly scaling with the size of the code, but since our design is flexible we study the effects of different trap capacities in Figure \ref{fig:sensitivity_tight}. For certain values of $x$ and the corresponding ion trap capacity for a given code, it is sometimes possible to achieve a lower overall execution time, depending on the physical parameters: $s$ (combined time for split, move, junction-cross, and merge operations), $t$ (swap time), $g$ (gate time), $x$ (number of traps), and the code-size-dependent parameter $m$. The base form of Cyclone is the sparsest configuration possible, but an ideal version of Cyclone balances all of these factors, yielding the most effective trap density. A space-efficient Cyclone model with $x$ traps yields a total worst case execution time of $2x \times (s + \lceil \frac{m}{x}\rceil\times(t + g\times\lceil \frac{xn}{m} \rceil))$.

In Figure \ref{fig:sensitivity_tight}, we examine precisely how Cyclone scales across many trap amounts (and therefore corresponding ion capacities) for a $[[225,9,6]]$ code and we find the ideal configuration to be a 64 trap architecture with a capacity of 8 ions per trap. It is worth noting that gate times $g$ scale very poorly after capacities greater than around $\sim$ 15 \cite{murali}, which heavily influence this outcome.

\subsection{Analysis of Experiments}

\begin{figure}[htbp!]
    \centering
    \includegraphics[width=0.95\linewidth]{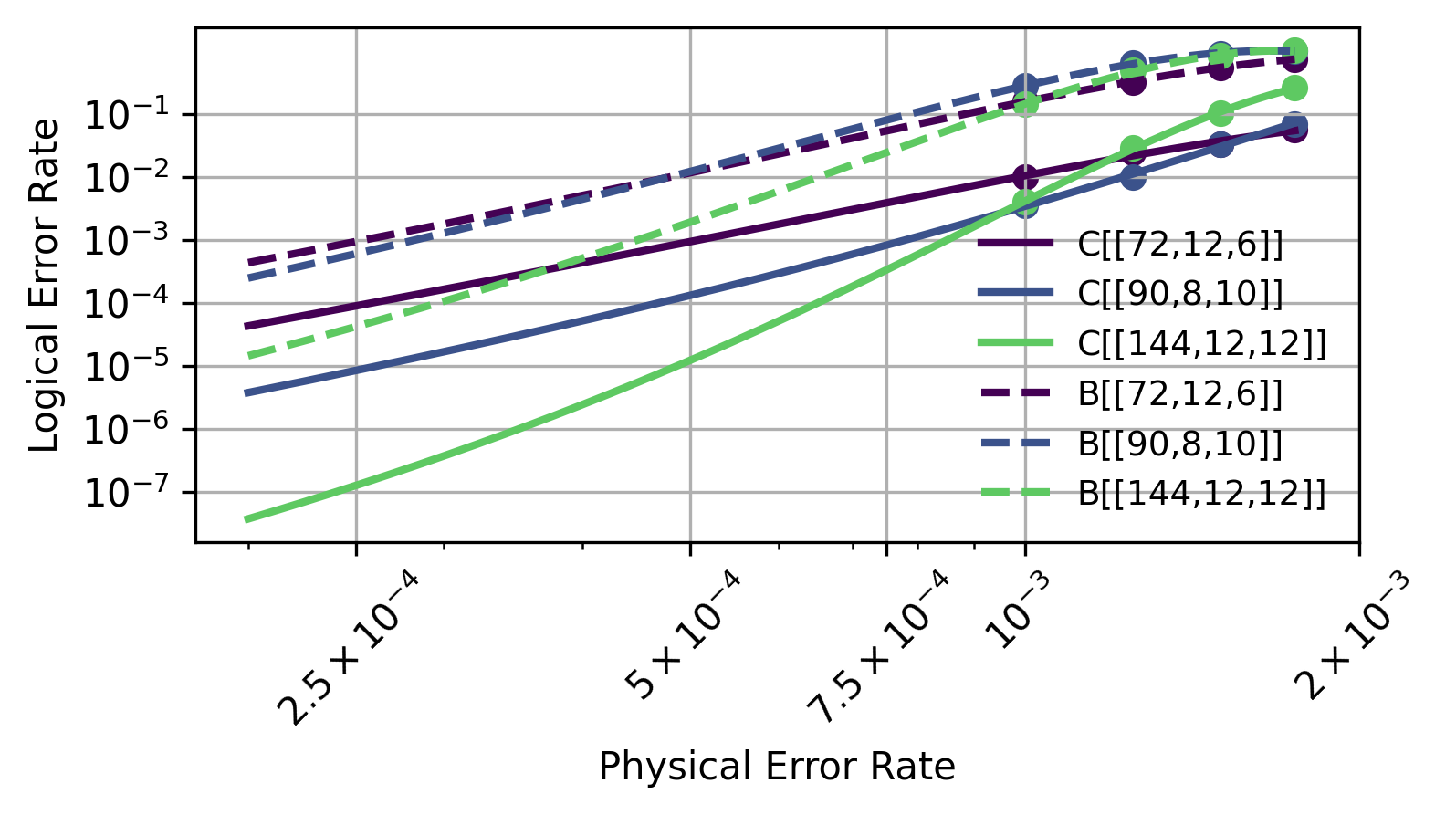}
    \caption{Comparing logical error rates between Cyclone (labeled as C) and the baseline (labeled as B) on different BB codes.}
    \label{fig:css_cyclone_ler}
\end{figure}

\begin{figure}[htbp!]
    \centering
    \includegraphics[width=0.95\linewidth]{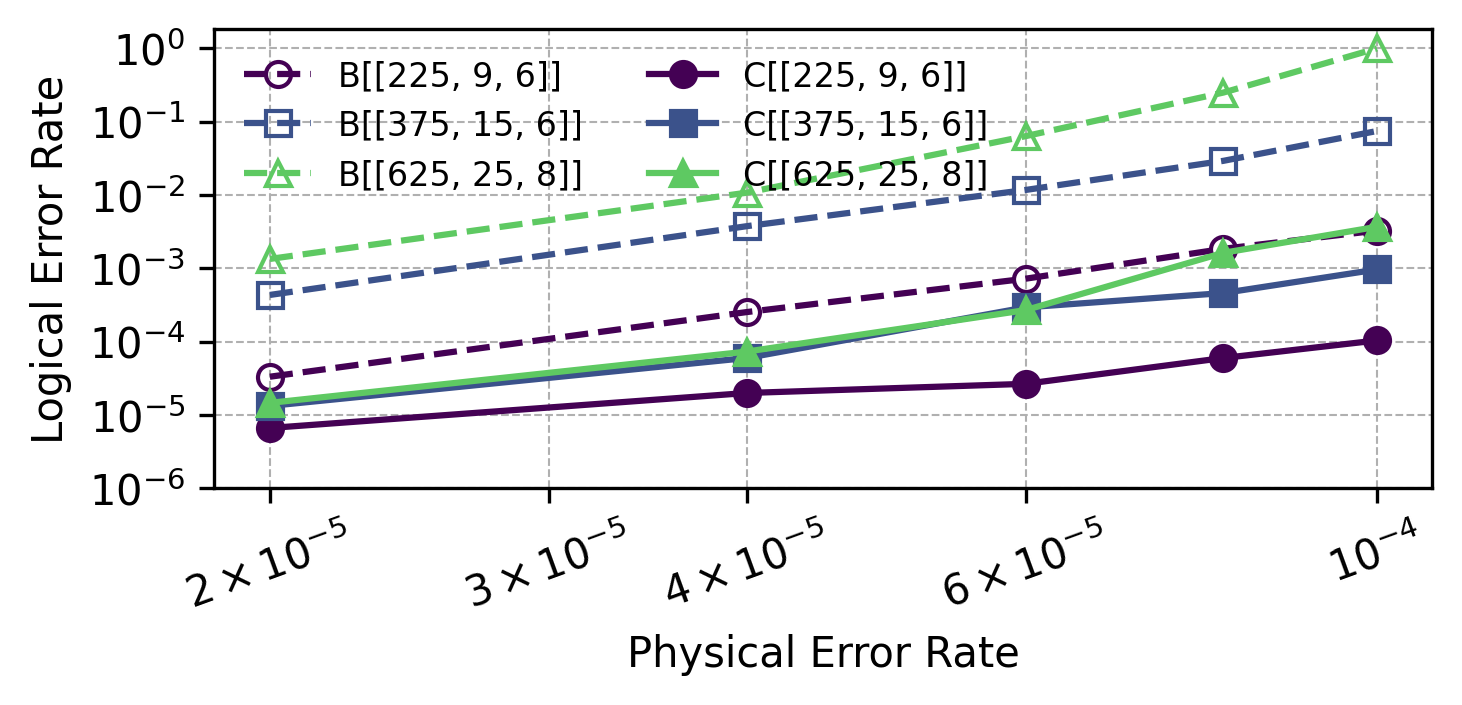}
    \caption{Comparing logical error rates between Cyclone (labeled as C) and the baseline (labeled as B) on different HGP codes.}
    \label{fig:qldpc_cyclone_ler}
\end{figure}

We observe a significant improvement in logical error rate (LER) with the base Cyclone codesign as opposed to the baseline. As circuit depth grows exponentially with increasing code size, the performance gap between Cyclone and baseline for larger and higher distance codes start to appear at progressively lower physical error rates. Moreover, the relative improvement in LER scales favorably with the code size. In the case of BB codes, differences emerge at higher physical error rates, and are larger in magnitude. For instance, the $[[144,12,12]]$ code exhibits up to 3 orders of magnitude improvement in the LER with Cyclone as shown in Figure \ref{fig:css_cyclone_ler}. In fact, only the baseline version of the $[[144,12,12]]$ code is even able to perform error correction within our tested physical error range, whereas in Cyclone error correction is seen for all codes. For HGP codes, Cyclone demonstrates two orders of magnitude in logical error rate improvement (Figure \ref{fig:qldpc_cyclone_ler}). Once again, error correction can be seen on all codes in the selected $p$ range with Cyclone, whereas the baseline fails to do so, unless sampled at lower $p$. These results suggest that Cyclone codesign will help enable error correction at higher levels of $p$, making large, high-rate non-topological CSS codes more practical for near-term trapped-ion architectures.

\begin{figure}[htbp!]
    \centering
    \includegraphics[width=0.95\linewidth]{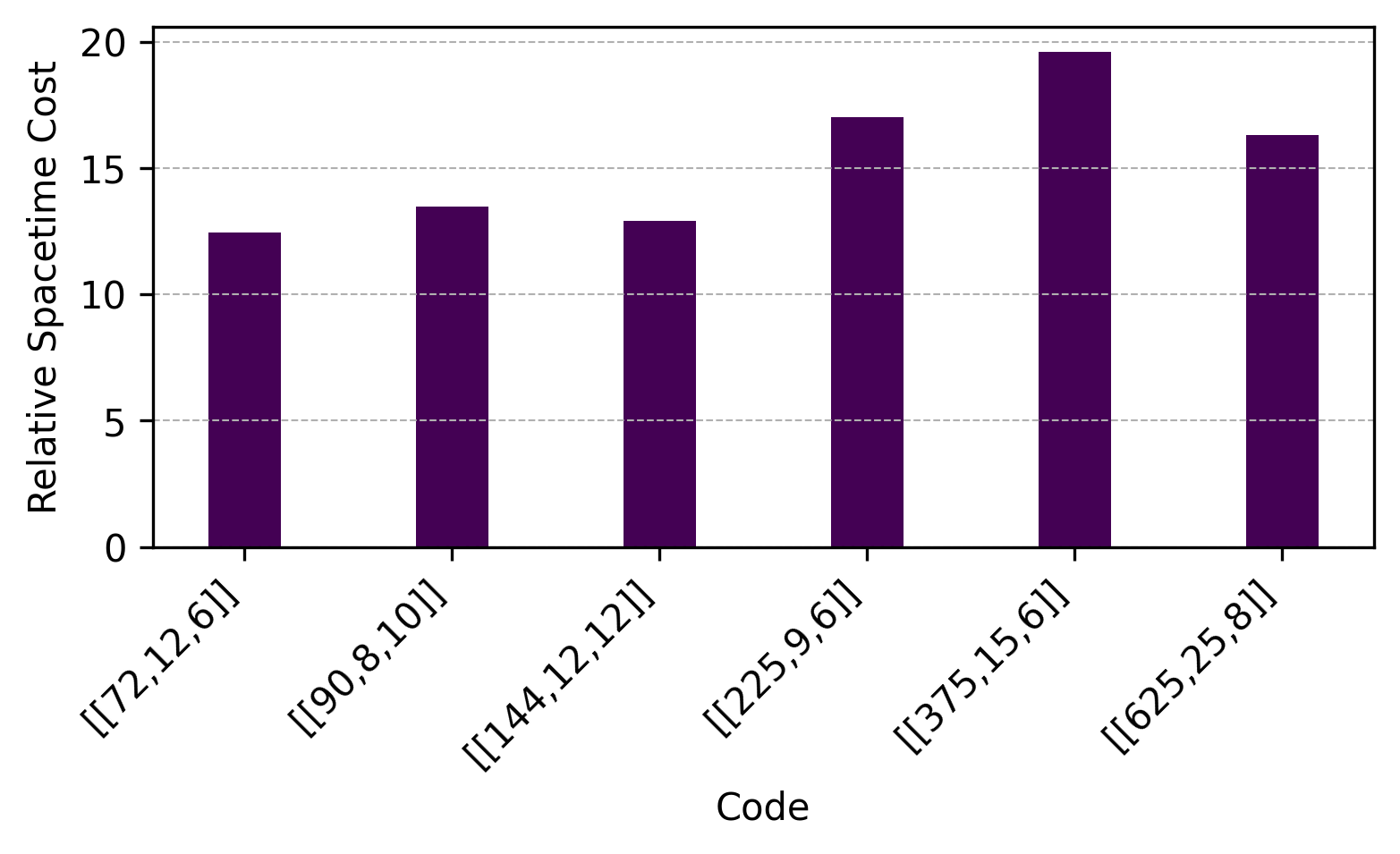}
    \caption{Spacetime cost of Baseline compared to relative spacetime cost of Cyclone. Spacetime is computed using Number of Traps $\times$ Execution Time $\times$ Number of Ancilla Qubits.}
    \label{fig:spacetime_cost}
\end{figure}

In Figure \ref{fig:spacetime_cost}, we analyze the relative spacetime cost of the baseline as compared to base Cyclone. Costs are computed by multiplying the number of traps by the execution time and the number of ancillary qubits. We attribute the differences in spacetime cost to the random performance of the baseline relative to Cyclone.

\subsection{Considering Independent or Concurrent Loops}

Since Cyclone introduces some redundancy by requiring ancilla qubits to traverse a single global loop, we briefly investigate whether separating independent or concurrent/overlapping loops could provide more direct paths for certain stabilizers, similar to the approach used in Pseudo-OPT. If a separate loop is designated for a set of stabilizers and executed in parallel with the smaller global loop, some stabilizers will still share data qubits between the two loops, forcing the ancilla qubits to traverse both. Such separation is only feasible in local topological codes, such as the surface code. However, our analysis shows that neither HGP nor BB codes permit such cuts due to their long-range and non-local connections. Therefore, creating separate loops consistently performs worse than maintaining a single global loop, as it introduces additional shuttling operations, increases spatial overhead, and raises the likelihood of roadblocks, ultimately negating any potential benefits from parallelism. For these reasons, we retain Cyclone’s single-loop architecture.

\subsection{Sensitivity Analyses}
\label{sec:sensitivity}
We conduct a series of sensitivity studies to better understand Cyclone’s trade-offs. In Figure \ref{fig:sensitivity_tight}, we explore the sensitivity to different plausible configurations that have minimal excess space we call ``tight" architectures. In these architectures, the trap ion capacity is the minimum amount needed to fit the code on the number of traps $x$. The formula is split for ancilla and data, so the capacity was chosen as $\lceil\frac{225}{x}\rceil + \lceil\frac{216}{x}\rceil$, where 225 and 216 are the respective data and ancilla counts for the chosen $[[225,9,6]]$ code in this analysis. This allows us to explore how different trap counts and capacities affect performance, and compare performance relative to the baseline efficiently. In the case of one trap and 441 ions, these architectures are equivalent with no shuttling. Designs with dense capacity and a large number of traps not only have more serialization, but the gate time scales as a function of trap capacity, typically resulting in long execution times for large codes. We find that even with as few as 9 traps, Cyclone significantly outperforms the baseline grid codesign, highlighting its superior parallelism and scheduling efficiency.

\begin{figure}[htbp!]
    \centering
    \includegraphics[width=0.95\linewidth]{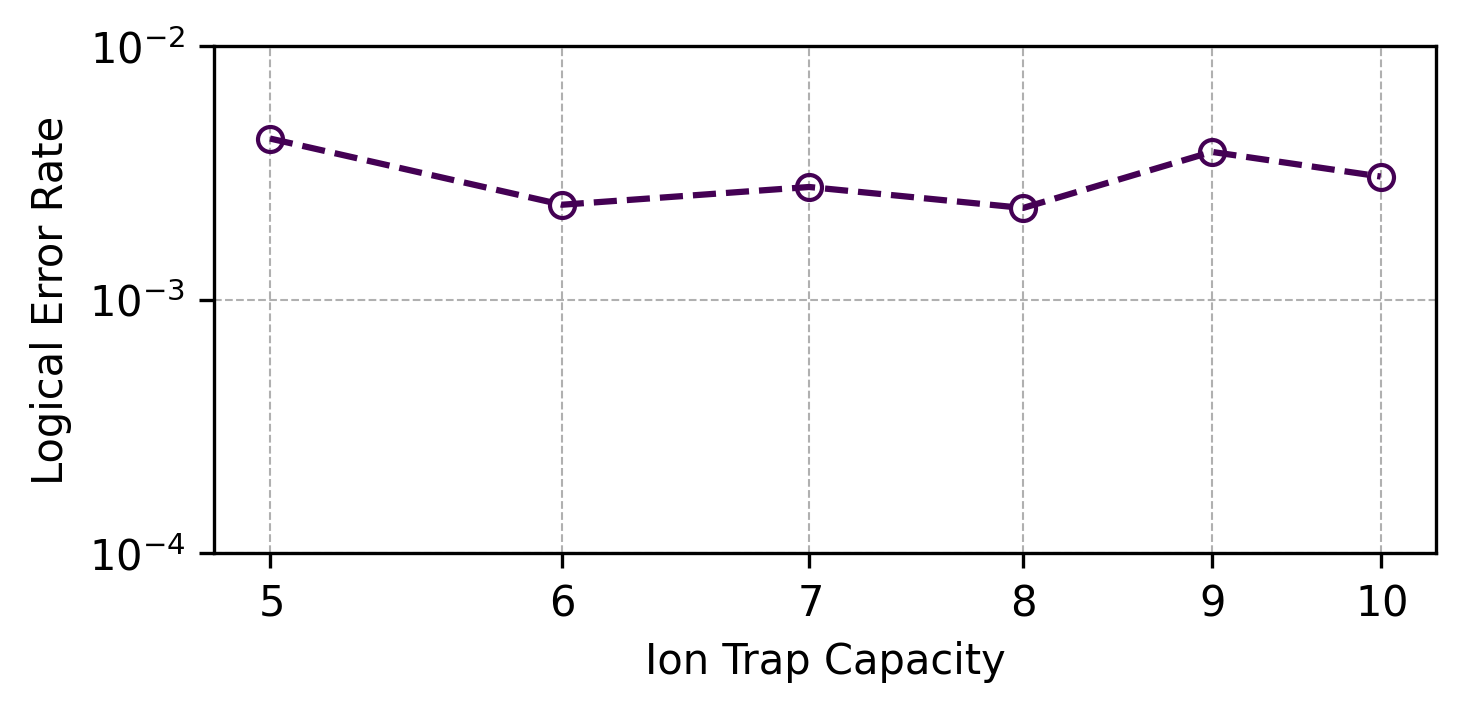}
    \caption{Comparing Sensitivity of the Baseline LER to a loosely fitting trap arrangement that has excess capacity at a constant physical error rate of $10^{-4}$ on a $[[225,9,6]]$ code}
    \label{fig:sensitivity_loose}
\end{figure}

We also explore sensitivity to ion capacity on ``loosely" fitting designs (as seen in Figure \ref{fig:sensitivity_loose}), where extra space is given to the baseline in order to check if architectural ``tightness" was constraining performance. In this example, the baseline ion capacity value used for previous experiments was 5, and we see negligible performance improvement upon allowing larger trap capacities, ensuring that the baseline is not architecturally constrained by the number of traps nor the capacity of the machine. 

\begin{figure}[htbp!]
    \centering
    \includegraphics[width=0.85\linewidth]{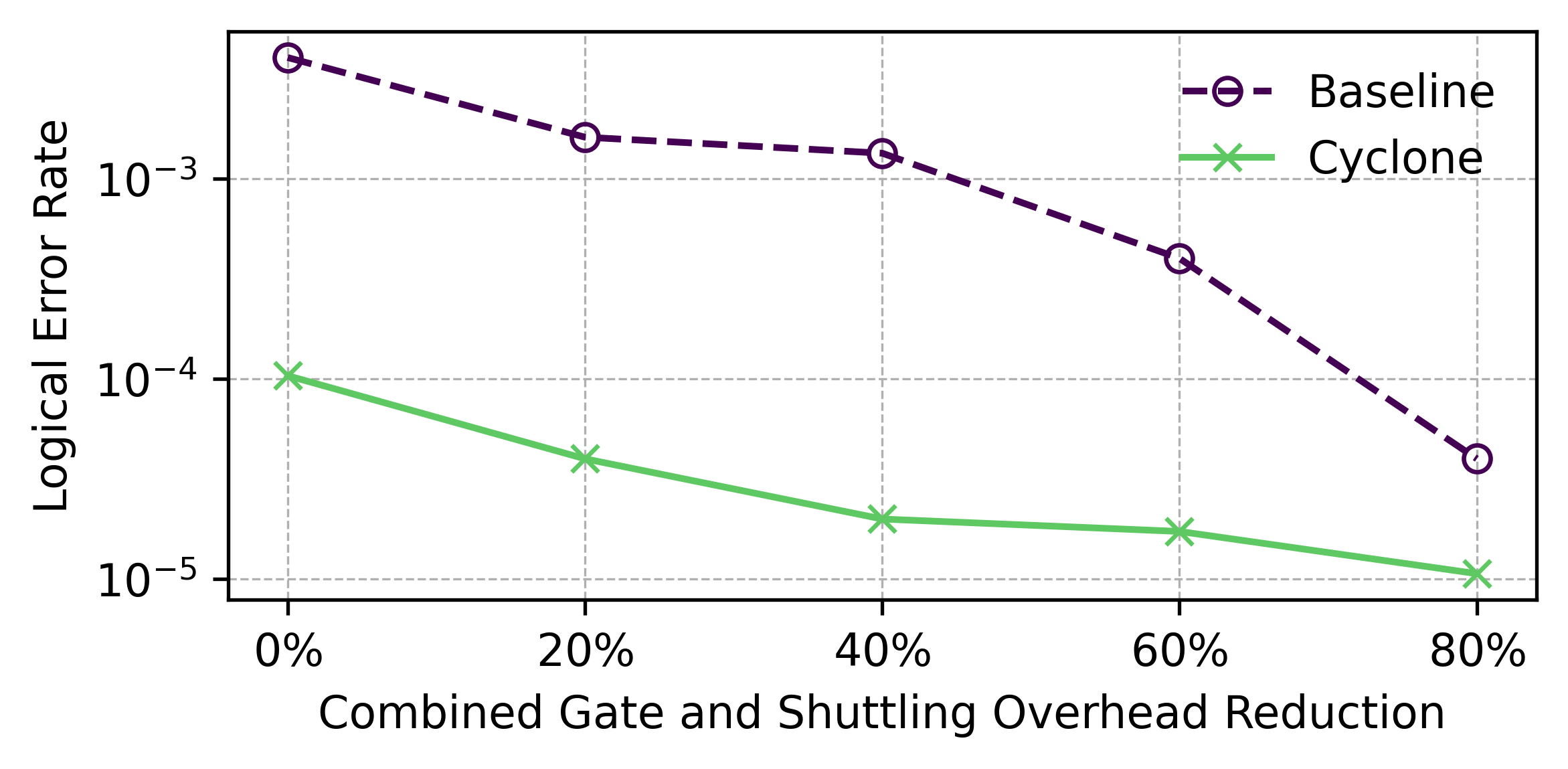}
    \caption{Sensitivity analysis to reducing both shuttling and gate times x\% on a $[[225,9,6]]$ code. As these times improve, the logical error rate becomes more limited based on the error-correcting ability of the code.}
    \label{fig:sensitivity_shuttling}
\end{figure}

Furthermore, we analyze the sensitivity of performance to improvements in gate and shuttling (split, move, merge, and junction crossing) times, by uniformly reducing their durations by a fixed percentage, $r$, as shown in Figure \ref{fig:sensitivity_shuttling}. In this experiment, we assume a physical error rate of $10^{-4}$ and use a $[[225,9,6]]$ HGP code. As expected, the performance gap between the baseline and Cyclone narrows, as improvements in operation time are offset by the error-correction capability of the code, which becomes the limiting factor in reducing the logical error rate, as $r$ increases.

\begin{figure}[htbp!]
    \centering
    \includegraphics[width=0.95\linewidth]{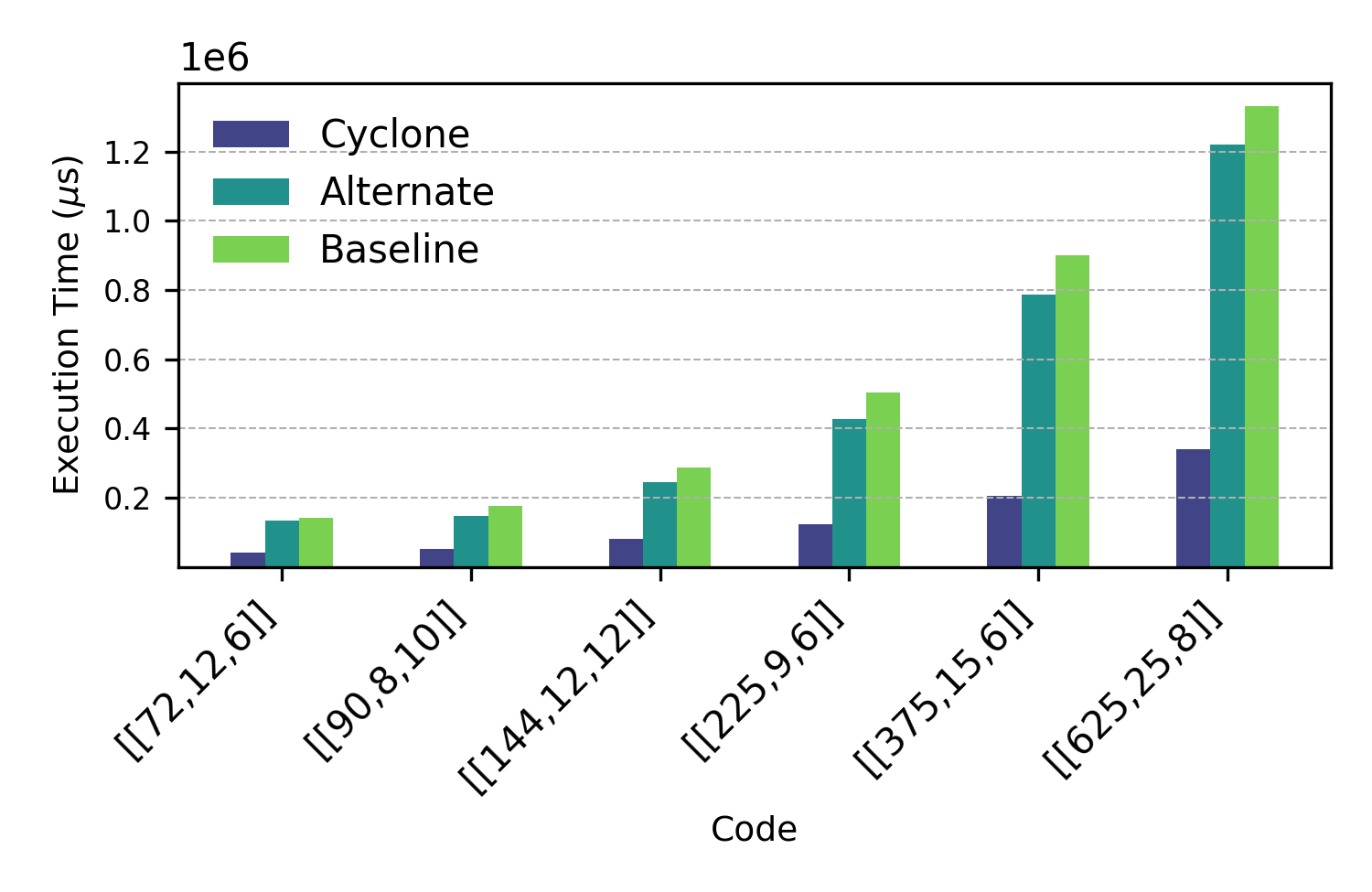}
    \caption{Sensitivity Analysis of execution times when accounting for the alternate grid, baseline, and Cyclone architectures. Raw execution times are compared as opposed to logical error rate.}
    \label{fig:sensitivity_grids}
\end{figure}

\begin{figure}[htbp!]
    \centering
    \includegraphics[width=0.95\linewidth]{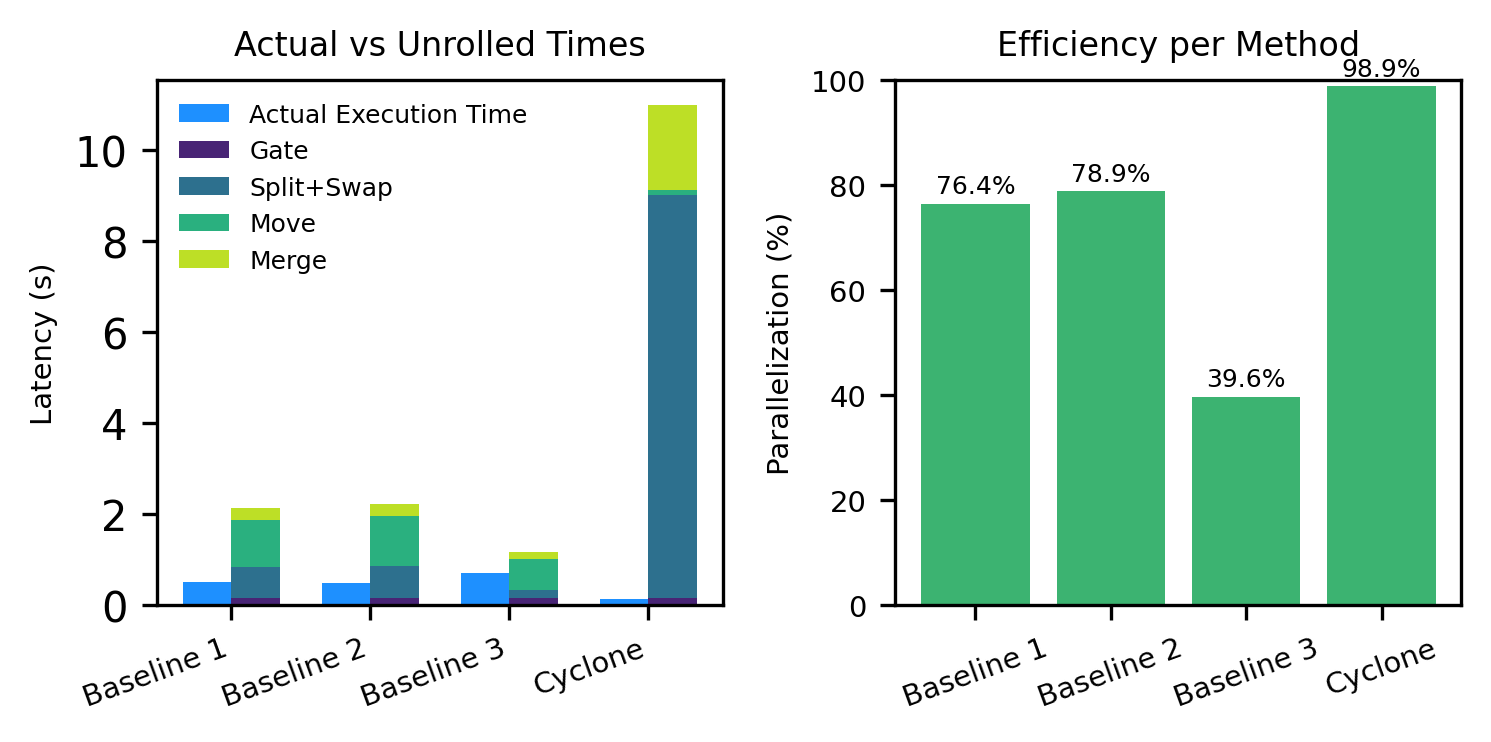}
    \caption{Left: A plot showing the total execution times and adjacent to it the \textit{unrolled} component-wise execution times on three different baseline compilers given the same architectural setup as detailed in Section \ref{sec:evaluation}\cite{murali}\cite{10.5555/3539845.3539926}\cite{khan2025movelessminimizingoverheadqccds}. Right: The corresponding \% parallelization found by taking the fraction of the actual execution time divided by the serialized components.}
    \label{fig:component_times}
\end{figure}

\begin{figure}[htbp!]
    \centering
    \includegraphics[width=0.85\linewidth]{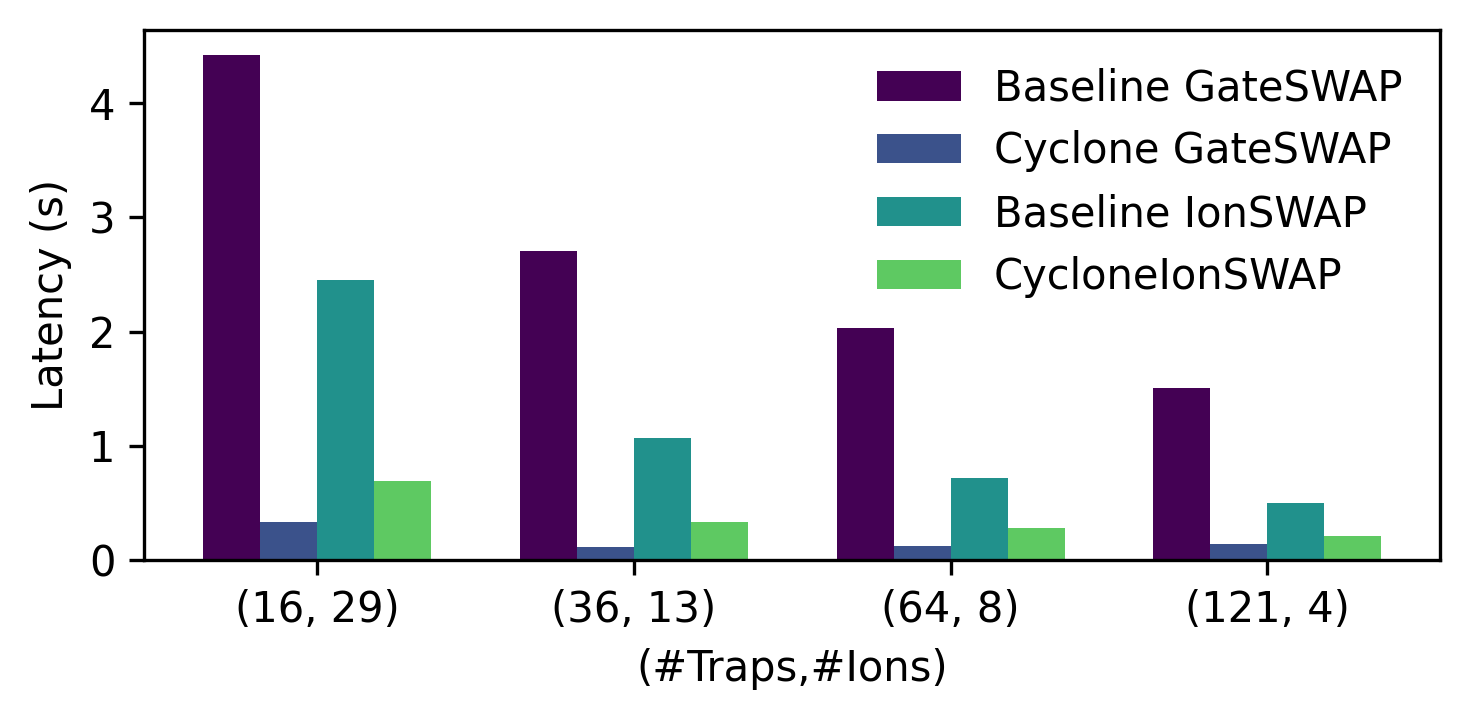}
    \caption{Sensitivity analysis to IonSWAP vs GateSWAP. IonSWAP scales with the interaction distance of the candidate ions, and GateSWAP is a function of gate time (which is constant for chain length 12 and under).}
    \label{fig:sensitivity_ion_swap}
\end{figure}

We explore sensitivity to the hardware: we change the type of grid design used to make our baseline to the alternate previously shown in Figure \ref{fig:background-grids}. We find that for the HGP and BB codes, an alternating vertical/horizontal mesh performs better than the standard grid we called baseline \cite{murali} (as opposed to allowing full vertical transport through junctions). Still, Cyclone outperforms the baseline and alternate grid by a wide margin as seen in Figure  \ref{fig:sensitivity_grids}. Execution times are used as the metric for clarity, as it exposes differences between the alternate grid + static EJF scheduling and the baseline more clearly. Similarly, in Figure \ref{fig:component_times}, we evaluate compiler sensitivity by comparing two additional baseline compilers—Baseline 2 \cite{Saki_2022} and Baseline 3 \cite{khan2025movelessminimizingoverheadqccds}—under the same architectural setup as the original baseline. We present both the component-wise breakdown of the fully serialized (unrolled) execution times and the realized total execution time. While most compilers achieve reasonable parallelization, Cyclone demonstrates a significantly more coordinated and highly parallel schedule. We attribute this improvement to Cyclone’s ability to perform O(m) operations per time step, whereas grid-based roadblocks inherently limit achievable parallelism.

Finally, we study sensitivity to the choice to use GateSWAP over IonSWAP as our swap technique throughout our paper (Figure 
\ref{fig:sensitivity_ion_swap}) Our decision of using GateSWAP was motivated by the fact that IonSWAP scales more directly with the size of the trap ($3 \times g$), while IonSWAP is $s \times d_l + s \times (d_l-1) + 42$ where $d_l$ is the interaction distance from the end to the ion \cite{murali}. We find that in general the Baseline performs better using IonSWAP and Cyclone using GateSWAP, and that Cyclone still keeps a convincing speedup regardless.

\section{Simulator Framework and Evaluation Methodologies}
\subsection{Hardware Simulation}
\label{sec:evaluation}

We conducted our simulations using the open source simulator QCCDSim~\cite{murali}. The quantum error correcting codes used in our study include Hypergraph Product codes found in \cite{quits}, and Bivariate Bicycle codes as found in \cite{ibmBB}. We use the default gate time models, shuttling times (as detailed in \ref{sec:shuttling-overhead}), and frequency modulated gates in our simulations. The baseline software simulator simulation policy schedules according to an earliest job first approach based on the program's interaction DAG. Alternatively, the improved dynamic software policy scraps the DAG representation and schedules according to the appropriate maximal parallelization policy in a given time step (found in Section \ref{sec:finding-ideal-parallel}), disregarding the more restrictive dependent gate scheduling. The baseline hardware is a square $l \times l$ grid where $l = \lceil \sqrt{n} \rceil$, and $n$ is the number of data qubits in the code. This should theoretically allow for the maximum amount of parallelism for baseline, and sensitivity to different ion capacities of the baseline are studied in Section \ref{sec:sensitivity}.   

Cyclone uses a custom compiler using the same operation times as in \cite{murali} to build a precise schedule of the atomic steps, and outputs a final execution time and schedule.

\subsection{Memory Experiments}

We use all of the aforementioned execution times and the corresponding QEC codes to conduct hardware-aware memory experiments. We input the latency to create realistic noise models as detailed in Section \ref{sec:noise-model}. We use the decoder for bivariate bicycle codes \cite{ibmBB}, and the QuITS decoder for HGP codes \cite{quits} all with error $p = p_{base} + p_{twirling}$ \cite{paulitwirl1, paulitwirl2}. We sample the decoders with enough shots $>10 \times$ $\frac{1}{L}$, where $L$ is the logical error rate. We assume sympathetic cooling can be implemented after every gate operation and mitigate most of the effects of heat accumulation \cite{sympathetic_cooling}.

 \begin{figure}[htbp!]
    \centering
    \includegraphics[width=0.85\linewidth]{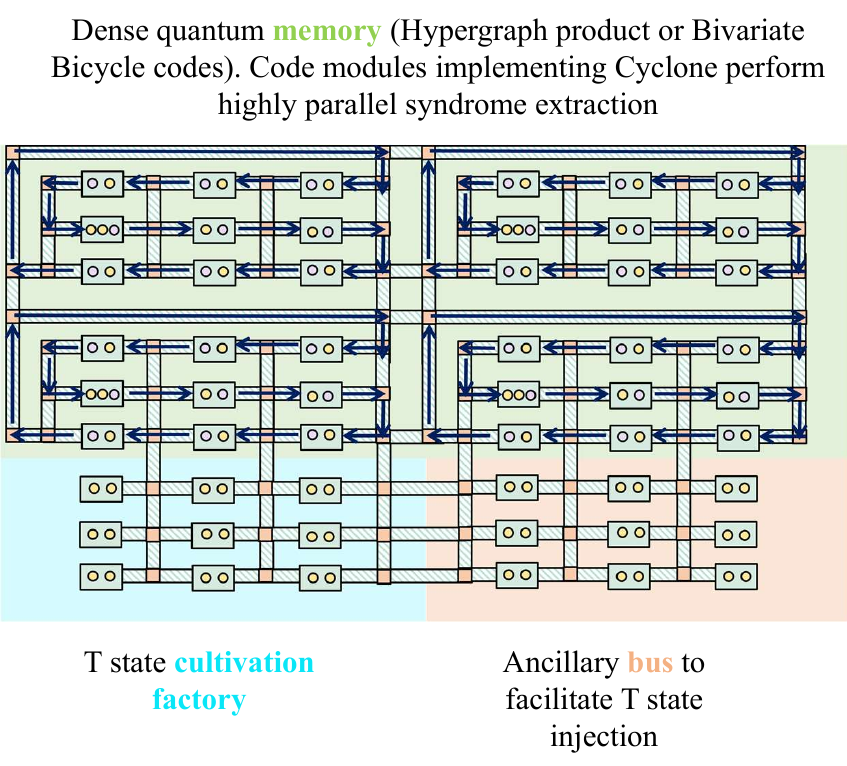}
    \caption{A feasible implementation of how Cyclone on qLDPC creates specific grids in the context of a universal fault-tolerant trapped ion quantum computer. An ancillary bus can be used to facilitate logical operations and T injection from the T state cultivation factory.}
    \label{fig:discuss_fig}
\end{figure}

\section{Discussion}

We believe that Cyclone will impact the way dense quantum memories are designed for fault tolerant quantum architectures on trapped ions in two possible directions. First, one could wish to preserve the generalized grid structure due to the fact that grids are generally a better use of space on a chip and allow flexibility for more possible movements. In this case, one can make slight but \textbf{purposeful} alterations to industry roadmaps \cite{Dac-forwarding}\cite{oxfordionicsgrid} to allow for loops as shown previously in Figure \ref{fig:cyclone_base_design}b. Another possibility is that since grids could be complex to design and wire~\cite{DAC_savings}, quantum memory in the shape of a ring in could be simpler to design. Cyclone allows this architectural flexibility while still maintaining significant improvement over state of the art. Figure \ref{fig:discuss_fig} shows how we envision a feasible practical implementation of Cyclone embedded in a universal fault-tolerant architecture, with a cultivation factory and an ancillary bus to facilitate T injection. In the figure, we have shown how one can combine four code blocks together to make an arbitrary-sized dense quantum memory. In the setup shown, logical Clifford operations can be implemented by using modified stabilizer measurements, or expanding the size of the ancilla bus to facilitate qLDPC surgery. Implementation of efficient logical operations and T injection from distillation/cultivation contains many different design choices, each with their own ensuing tradeoffs. We leave the exploration of these choices and tradeoffs as a natural follow up for future work.

\section{Conclusion}

We find that standard grid architectures from industry (those lacking topological loops) are insufficient for non-topological CSS codes due to their inability to support high levels of parallelism. The emerging high-rate Bivariate Bicycle and Hypergraph Product codes are examples of non-topological CSS codes that offer highly parallel syndrome extraction circuits, leaving plenty of room for hardware and software optimization. We search through the design space by reverse engineering from an idealized, fully connected layout and present strategies to make those designs practical. We observe that many designs require high connectivity, rely on the quality and practicality of crossing many junctions on a chip, which at the moment, is not practical. 

Our exploration motivates a symmetric software policy along with a circular hardware design that has no roadblocks and a one-directional path we call \textit{Cyclone}. Cyclone achieves up to $4 \times$ speedup, amounting to up to $2-3$ orders of magnitude improvements in logical error rates. In addition to the temporal improvements, Cyclone is also spatially efficient, using fewer traps than a grid and typically allowing for denser traps in its optimal form, saving on resources. Cyclone also requires a constant number of Digital-to-Analog Converters (DACs) as a single control signal that can be broadcast to all the traps, significantly alleviating the qubit wiring problem detailed in \cite{DAC_savings, Dac-forwarding}. We envision Cyclone will reshape how dense memory is constructed in fault tolerant trapped ion architectures. Furthermore, Cyclone's design principle may extend to silicon quantum dots, which share similar architectural principles to QCCDs.

%%%%%%% -- PAPER CONTENT ENDS -- %%%%%%%%

%%%%%%%%% -- BIB STYLE AND FILE -- %%%%%%%%
\bibliographystyle{IEEEtranS}
\bibliography{refs}
%%%%%%%%%%%%%%%%%%%%%%%%%%%%%%%%%%%%

\end{document}